\documentclass[12pt]{article}
\usepackage{cite}
\usepackage[T1]{fontenc}
\usepackage[utf8]{inputenc}

\usepackage[english]{babel}

\usepackage{graphicx}
\usepackage[dvipsnames]{xcolor}
\usepackage{float}

\usepackage{amsmath}
\usepackage{amsfonts}
\usepackage{amssymb}
\usepackage{amstext}
\usepackage{slashed}
\usepackage{braket}
\usepackage{caption}
\usepackage{subcaption}
\usepackage{layouts}
\usepackage[normalem]{ulem}
\usepackage{physics}

\usepackage[bookmarks, breaklinks, colorlinks,urlcolor=black, citecolor=red, linkcolor=blue]{hyperref}

\usepackage{array}

\def\dsp{\displaystyle}
\def\bea{\begin{align}}
\def\eea{\end{align}} 
\def\be{\begin{equation}}
\def\ee{\end{equation}} 
\def\nn{\nonumber}

\def\tev{\ensuremath{\mathrm{Te\kern -0.1em V}}}
\def\gev{\ensuremath{\mathrm{Ge\kern -0.1em V}}}
\def\mev{\ensuremath{\mathrm{Me\kern -0.1em V}}}
\def\mdm{\ensuremath{m_\text{DM}}}
\def\gdm{\ensuremath{g_\text{DM}}}

\renewcommand{\d}{\operatorname{d}\!}

\newcommand{\VEV}[1]{\langle #1 \rangle}

\def\omg{\ensuremath{\Omega h^2}}
\def\sv{{\langle \sigma v \rangle}}
\def\se{s_\text{ent}}
\def\ecm{E_\text{CM}}
\def\TR{T_\text{RH}}
\def\YIR{Y_0^{\text{IR}}}
\def\YUV{Y_0^{\text{UV}}}
\def\gff{\ensuremath(g_\pm^f)_{ij}}

\begin{document}
	
\begin{flushright}
SI-HEP-2023-18
\end{flushright}

\vskip 2cm	
	
	\begin{center}
		
		{\Large\bf Exploring Freeze-out and Freeze-in Dark Matter \\[0.8ex] via Effective Froggatt-Nielsen Theory} \\[8mm]
		{Rusa Mandal \footnote{Email: rusa.mandal@iitgn.ac.in} and Tom Tong \footnote{Email: Tiantian.Tong@uni-siegen.de}}

  {\small\em $^1$Indian Institute of Technology Gandhinagar, Department of Physics, \\ Gujarat 382355, India}

    {\small\em $^2$Center for Particle Physics Siegen (CPPS), Theoretische Physik 1,  \\Universit\"at Siegen, 57068 Siegen, Germany}

	\end{center}

\begin{abstract}

Motivated by the dynamical reasons for the hierarchical structure of the Yukawa sector of the Standard Model (SM), we consider an extension of the SM with a complex scalar field, known as `flavon', based on the Froggatt-Nielsen mechanism. In an effective theory approach, the SM fermion masses and mixing patterns are generated in orders of the parameter related to the vacuum expectation value of the flavon field and the cut-off of the effective theory. By introducing right-handed neutrinos, we study the viability of the lightest right-handed neutrino as a dark matter candidate, where the same flavon field acts as a mediator between the dark and the SM sectors. We find that dark matter genesis is achieved both through freeze-out and freeze-in mechanisms encompassing the $\mathcal{O}(\gev)$ -- $ \mathcal{O}(\tev)$ mass range of the mediator and the dark matter particle. In addition to tree-level spin-dependent cross section, the model gives rise to tree- and loop-level contributions to spin-independent scattering cross section at the direct detection experiments such as XENON and LUX-ZEPLIN which can be probed in their future upgrades. By choosing suitable Froggatt-Nielsen charges for the fermions, we also generate the mass spectrum of the SM neutrinos via the Type-I seesaw mechanism. Flavor-changing neutral current processes, such as radiative lepton decay, meson mixing, and top-quark decay remain the most constraining channels and provide testability for this minimal setup that addresses several major shortcomings of the SM.
  
\end{abstract}

\newpage
{\hypersetup{linkcolor=black}\tableofcontents}
\section{Introduction}

Unraveling the hierarchical fermion mass structure remains a puzzle within the framework of the Standard Model (SM) of particle physics. This arrangement of the Yukawa sector of the SM can potentially be attributed to new symmetries that unify the lepton and quark sectors. A popular setup to explain this hierarchical structure is the Froggatt-Nielsen (FN) mechanism~\cite{Froggatt:1978nt}. In this mechanism, the SM Yukawa structure is dynamically generated through the spontaneous breaking of a $U(1)$ symmetry by a scalar field known as the `flavon,' which acquires a vacuum expectation value ({\it vev}). All SM fermions are charged under that new symmetry, leading to an interaction of the flavon field with them and thus enhancing the detectability of the flavon field.  Interestingly, the pseudoscalar component of the flavon field can serve as a QCD axion field. This arises when the global FN symmetry is identified with the global Peccei-Quinn symmetry~\cite{Peccei:1977hh}, introduced to solve the strong $CP$ problem~\cite{Weinberg:1977ma}. 
However, in the absence of such an identification, pseudoscalar flavons can populate a heavier mass range and exhibit rich phenomenology. Some studies in these directions can be found in~\cite{Ema:2016ops,Calibbi:2016hwq,Bauer:2016rxs}.

Some of the demanding reasons for extending the SM include the observation of nonzero neutrino masses and astrophysical evidence of dark matter (DM). In this work, we combine these two motivations with the dynamical reasons for the hierarchical flavor structure. We used flavon as a portal between the dark sector and the SM where the dark sector consists of right-handed neutrinos (RHNs). The feasibility of RHNs as DM candidates within the framework of FN symmetry has been discussed here~\cite{Merle:2011yv,Jaramillo:2020dde}. We extended the SM with three generations of heavy RHNs, where the lightest of them is a candidate for DM, and the others are responsible for the generation of neutrino masses by the Type I seesaw mechanism.

The model setup considered in this work is economic and spans a quite wide range of the mass spectrum. Interestingly, the interaction strength of the flavon with the DM candidate can be such that the DM may or may not be in thermal equilibrium with the SM. Therefore, the genesis of DM is realized separately through both freeze-out and freeze-in mechanisms. In such a flavon model as the SM fermion mass hierarchy is achieved in powers of small parameter proportional to the {\it vev} of the flavon field divided by the cutoff of the theory, the model parameter space is very restrictive. This in turn increases the predictive power of the theory. We find that as the strength of interaction between new physics and SM particles is constrained by the observed mass and mixing spectrum, the single parameter, that is, the {\it vev} of the flavon field $v_\phi$, mainly determines the nature of the genesis of DM to be achieved.  

Our analysis focuses on relatively heavy mediators and the DM mass spectrum compared to some previous analyses where the freeze-out mechanisms~\cite{Calibbi:2015sfa} and freeze-in mechanisms~\cite{Cheek:2022yof,Babu:2023zni} are studied separately through the scalar flavon portal. Flavon theories inherently induce the flavor-changing neutral current (FCNC) at the tree level. From the outcomes of current experimental searches in both quark- and lepton-level FCNC processes, the FN symmetry breaking scale requires to be advanced at least to the TeV range. In light of a naturalness perspective, the core of these theories rely on the dynamical generation of mass hierarchy which avoids small bare coupling parameters at the Lagrangian level. This makes the mass of the scalar component of the flavon at the FN symmetry breaking scale, whereas the mass term for the pseudoscalar part (the pseudo-Goldstone boson of spontaneous breaking of FN symmetry) arises only from the soft-breaking term, which is supposed to be much smaller than its scalar counterpart. In the absence of such a soft-breaking term, a massless Goldstone boson featuring flavor-violating couplings does not align favorably with the ongoing experimental searches or astrophysical observations conducted over the past few decades~\cite{Kim:2008hd,Jaeckel:2010ni}. We focus mainly on the pseudoscalar portal due to the prospects of discovery in recent experiments. We explore both the freeze-out and freeze-in mechanisms for the DM productions and identify the parameter space explaining the observed relic density of the Universe. Our findings show that for a mediator $\mathcal{O}(\gev)$, the freeze-out mechanism works with the DM candidate in the $\mathcal{O}(\tev)$ range, and the framework provides spin-independent and spin-dependent scattering cross sections in direct detection experiments. However, for freeze-in production of a DM below 1 TeV, a significantly higher value of $v_\phi$, $\mathcal{O}(10^7-10^9)\,$GeV is required. The choice of reheating temperature of the Universe also plays an important role in identifying the processes contributing to the DM production. It is worth mentioning that the entire framework can produce the observed mass and mixing texture for all SM fermions, including non-zero masses for light neutrinos. 

The paper is organized as follows. In Sec.~\ref{sec:model}, we define the low-energy effective Lagrangian generated after spontaneous breaking of the FN symmetry. The interactions between the flavon and the SM along with the RHN fields are worked out. Section \ref{sec:constraint} deals with listing the stringent experimental constraints arising from both low-energy processes such as meson mixing (in \ref{sec:Meson}), radiative lepton decay (in \ref{sec:muegamma}), and direct searches (in \ref{sec:Top}). We then discuss the scenarios to generate the observed relic abundance in Sec.~\ref{sec:freeze-out} and in Sec.~\ref{sec:freeze-in} via freeze-out and freeze-in mechanisms, respectively. Section~\ref{sec:detection} deals with computing the spin-independent and spin-dependent scattering cross sections for direct detection experiments. We dedicate Sec.~\ref{sec:neutrino} to explicitly showing the framework for generating light neutrino masses using the seesaw mechanism. The FN charge assignments are worked out separately for the freeze-out and freeze-in cases. Finally, we summarize in Sec.~\ref{sec:summary} with some discussion. A benchmark example of the $\mathcal{O}(1)$ entries in the Yukawa matrices is given in Appendix~\ref{app:yukawa}. The relevant expressions for the amplitudes and cross sections that contribute to the annihilation processes are given in Appendix~\ref{app:Xsec}.

\section{Model}
\label{sec:model}

We start with a complex scalar flavon field $\phi$, which is singlet under the SM gauge group. An extra abelian $U(1)$ symmetry, known as the FN symmetry, is invoked where all SM fermions poses a distinguishable charge. The spontaneous breaking of the FN symmetry, through {\it vev} of $\phi$, is communicated to the fermions at different orders in a small parameter $\epsilon\equiv \langle \phi \rangle/M$. Here $M$ is the scale of flavor dynamics and is associated with some heavy fermions that are integrated out. Although the full theory will have many heavy fermions, called FN fields, the effective theory below $M$ is quite simple, and the interaction Lagrangian of the flavon field $\phi$ with SM fermions and three SM gauge singlet right-handed neutrinos $N_R^i$ can be written as
\begin{align}
\label{eq:Lag}
	-\mathcal{L}_\text{int} &=c^{i j}_{d}\left(\frac{\phi}{M}\right)^{n^{i j}_{d}} \bar{Q}^{i} H d_{R}^j+c^{i j}_{u}\left(\frac{\phi}{M}\right)^{n^{i j}_{u}} \bar{Q}^{i}\;i \sigma_2H^* u_{R }^j +c^{i j}_{e}\left(\frac{\phi}{M}\right)^{n^{i j}_{e}} \bar{L}^{i} H e_{R }^j \nn \\
	&+c^{ij}_{\nu}\left(\frac{\phi}{M}\right)^{n^{i j}_{\nu}} \bar{L}^{i}\, i \sigma_2 H^* N_{R }^j +\frac{1}{2} c^{ij}_{N}\left(\frac{\phi}{M}\right)^{n^{ij}_{N}} M \overline{N_{R}^{c\,i}} N_{R}^j \ + \ \text{h.c.} \,,
\end{align}
where $c_x^{ij}$ are all $\mathcal{O}(1)$ numbers with $i,j=1,2,3$. By invoking an extra $Z_2$ symmetry, we consider the lightest RHN, $N^1$, as a viable candidate for DM that is odd under the $Z_2$ symmetry, implying $c^{i1}_{\nu}=0$ and $c^{1k}_{N} = c^{k1}_{N} = 0$, for $k=2,3$. Here the differences of the $U(1)_\text{FN}$ charges of the fermions are defined as
\begin{equation}
\begin{aligned}
	&n^{i j}_{u}\equiv q_{Q_{i}}-q_{u_{j}},~~n^{i j}_{d} \equiv q_{Q_{i}}-q_{d_{j}},~~ n^{i j}_{e} \equiv q_{L_{i}}-q_{e_{j}}, \\[1ex]
	& \quad \quad~~ n^{i k}_{\nu} \equiv q_{L_{i}}-q_{N_k},~~n^{i j}_{N} \equiv -q_{N_i}-q_{N_j} \,,
\label{eq:n-to-q}
\end{aligned}
\end{equation}
where all $n_f^{ij} \geq 0$. Therefore, it is also implied that $q_{N_i} \leq 0$.

Expanding the complex scalar $\phi$ and the SM Higgs doublet $H$ around their corresponding {\it vev} as
\begin{align}
\label{eq:vevs}
\phi=v_{\phi}+\frac{1}{\sqrt{2}}(s+i a), \quad H=\left(\begin{array}{c}
	0 \\
	v_\text{EW} + \dfrac{h}{\sqrt{2}}
\end{array}\right),
\end{align}
and the interaction terms of the (pseudo)scalars with the fermions are generated as 
\begin{align}
\label{eq:LagintLR}
-\mathcal{L}_\text{scalar} = \sum_{f=u,d,e}
\bigg[& m_{ij}^f \left(1 + \frac{h}{\sqrt{2}v_\text{EW}} \right)  + \frac{m_f^{ij} n_f^{ij} (s+i a)}{\sqrt{2}v_\phi}  \bigg] \bar{f}_L^i f_R^j \,,
\end{align}
where $v_\text{EW}=174\,$GeV.

In order to obtain the SM fermion masses and the CKM mixing pattern, a viable solution is known, where $\epsilon$ is identified with the Cabibbo angle \cite{Ema:2016ops,Bauer:2016rxs}
\begin{align}
\epsilon = \frac{v_\phi}{M} = \Big[ V_{CKM} \Big]_{12} \approx 0.23 \,,
\end{align}
and 
\begin{align}
	&n^{ij}_u=\begin{pmatrix}
		8 & 4 & 3 \\
		7 & 3 & 2 \\
		5 & 1 & 0 
	\end{pmatrix},~~
	n^{ij}_d=\begin{pmatrix}
		7 & 6 & 6 \\
		6 & 5 & 5 \\
		4 & 3 & 3 
	\end{pmatrix},~~
	n^{ij}_e=\begin{pmatrix}
		9 & 6 & 4 \\
		8 & 5 & 3 \\
		8 & 5 & 3 
	\end{pmatrix} \,.
	\label{eq:nFN}
\end{align}

The corresponding FN charges have to be chosen appropriately. One such possibility is
\begin{align}
\label{eq:FNsol}
\begin{gathered}
\left(\begin{array}{ccc}
q_{Q_{1}} & q_{Q_{2}} & q_{Q_{3}} \\
q_{u} & q_{c} & q_{t} \\
q_{d} & q_{s} & q_{b}
\end{array}\right)=\left(\begin{array}{ccc}
3 & 2 & 0 \\
-5 & -1 & 0 \\
-4 & -3 & -3
\end{array}\right),~~
\left(\begin{array}{ccc}
q_{L_{1}} & q_{L_{2}} & q_{L_{3}} \\
q_{e} & q_{\mu} & q_{\tau}
\end{array}\right)=\left(\begin{array}{ccc}
1 & 0 & 0 \\
-8 & -5 & -3
\end{array}\right) \,.
\end{gathered}
\end{align}
We leave the discussion of the FN charge assignment of the RHNs to Sec.~\ref{sec:neutrino}, where we study neutrino mass generation for our model. An example of the benchmark choice for $c_x^{ij}$ (in Eq.~\eqref{eq:Lag}) is given in Appendix~\ref{app:yukawa}.

The potential of the complex scalar $\phi$ can be written as 
\begin{align}
\label{eq:potential}
-\mathcal{L}_\text{potential} = -m_\phi^2 |\phi|^2 - \mu (\phi^2 + \phi^{*2}) + \lambda_\phi |\phi|^4 + \lambda_{\phi H} |\phi|^2 |H|^2\,,
\end{align}
where the interaction term with the SM Higgs doublet induces mixing between the real scalar components ($s$ and $h$), thus modifying the SM Higgs observables. Current constraints from LHC only allow for insignificant mixing~\cite{Babu:2023zni}. Therefore, we choose $ \lambda_{\phi H} \approx 0$ to focus on the pseudoscalar portal. 
Note that the {\it vev} of $\phi$ (with $v_\phi^2 \approx \frac12 m_\phi^2 / \lambda_\phi$)  breaks the $U(1)_\text{FN}$ symmetry which should give rise to a massless Goldstone boson. However, due to an explicit symmetry breaking term in Eq.~\eqref{eq:potential} via the $\mu$ parameter, a mass term for the pseudo-Goldstone boson $a$ is also generated which is expected to be small compared to the scalar $s$ since $\mu << m_\phi$. We have
\begin{align}
\label{eq:masses}
m_s = 2 \sqrt{\lambda_\phi} \, v_\phi \approx \sqrt{2} \, m_\phi \qquad \text{and} \qquad  m_a = 2 \sqrt{\mu} \,.
\end{align}

After EW symmetry breaking, the mass matrices for the quarks and charged leptons are found to be
$$m_f^{ij}= c_f^{ij} \epsilon^{n_f^{ij}} v_\text{EW}~~\text{with}~~f=u,d,e\,.$$
The mass matrices can be diagonalised using the biunitary transformation
$$ \left( U_L^{\dagger f}  m_f U_R^f \right)_{ij} = \ m^f_i \delta_{ij}\,. $$
Here we used the following unitary rotations for flavor-to-mass eigenstate transformations.
\begin{align}
f_L^i \to U_L^{ij} f_L^j, \qquad f_R^i \to U_R^{ij} f_R^j\,.
\end{align}
The leading dimension-four interaction terms for all (pseudo)scalar fields in this setup read
\begin{align}
\label{eq:gaf}
-\mathcal{L}_\text{scalar} = \sum_{f=u,d,e}
\bigg[& m_i^f \left(1 + \frac{h}{\sqrt{2}v_\text{EW}} \right) \bar{f}^i f^i  \nn \\
&+  i a \left( (g_+^f)_{ij} \bar{f}^i \gamma_5 f^j + (g_-^f)_{ij} \bar{f}^i f^j  \right) \nn \\
&+ s \left( (g_+^f)_{ij} \bar{f}^i  f^j + (g_-^{f})_{ij} \bar{f}^i \gamma_5 f^j  \right) \bigg]\,,
\end{align}
where the couplings are~\cite{Ema:2016ops}
\begin{align}
\label{eq:g+}
(g_+^f)_{ij} \ &= \ \frac{1}{2 \sqrt{2}} \left( U_L^{f \dagger} \hat{q}_Q U_L^f -  U_R^{f \dagger} \hat{q}_f U_R^f  \right)_{ij} \frac{m_j^f+ m_i^f}{v_\phi}\,, \\
\label{eq:g-}
(g_-^f)_{ij} \ &= \ \frac{1}{2 \sqrt{2} } \left( U_L^{f \dagger} \hat{q}_Q U_L^f +  U_R^{f \dagger} \hat{q}_f U_R^f  \right)_{ij} \frac{m_j^f- m_i^f}{v_\phi}\,.
\end{align}
Here, $ (\hat{q}_X)_{ij}= q_{X_i} \delta_{ij}$ are the diagonal matrices of the FN charges. Note that because of the presence of generation-dependent FN charges, the flavon couplings cannot be diagonalised simultaneously with the mass matrices. As a result, flavor-changing interactions are generated.

The DM candidate $N^1$ interacts only through the last term in Eq.~\eqref{eq:Lag} which can be written (using Eq.~\eqref{eq:vevs}) as
\begin{align}
\label{eq:gDM}
-\mathcal{L}_\text{RHN} \ &\supset \ \frac{1}{2} \, c_N^{11} \epsilon^{n_N^{11}} \left(1 + n_N^{11}\, \frac{s + ia}{\sqrt{2} v_\phi} \right) M \overline{N_{R}^{c\,1}} N_{R}^1 \ + \ \text{h.c.} \nn \\[2ex]
&= \frac{1}{2} \, \mdm \overline{N^1} N^1 + g_{\rm DM} \left( s \overline{N^1} N^1+ i a \overline{N^1} \gamma_5 N^1 \right).
\end{align}
In the last line, we have defined $m_\text{DM}=c_N^{11} \epsilon^{n_N^{11}}M$, which is the Majorana mass term of the DM candidate $N^1$, and 
\begin{align}
    \gdm = - \frac{q_{N_1} \, \mdm}{\sqrt{2} \, v_\phi}\,,
\label{eq:gDM}
\end{align}
is the interaction strength of the DM with the (pseudo)scalar. We can see that the DM particle interacts with the SM fermions via both $s$ and $a$, giving rise to the scalar and pseudoscalar portal, respectively. 

We note that apart from the interaction terms quoted in Eq.~\eqref{eq:gaf}, a term at leading order in the SM fermions and scalar fields, although dimension-five, can arise in this setup,
\begin{align}
\label{eq:Lagdim5}
-\mathcal{L}_\text5 = \sum_{f=u,d,e} \frac{(s+ia)\,h}{\sqrt{2}\,v_\text{EW} } \left( (g_+^f)_{ij} \bar{f}^i \gamma_5 f^j + (g_-^{f})_{ij} \bar{f}^i f^j  \right)\,, 
\end{align}
which is further suppressed by the electroweak scale. This gives rise to a contact interaction between the fermions, the Higgs boson, and the (pseudo)scalar flavon. Later in Sec.~\ref{sec:freeze-in} we will discuss the impact of such a term on the freeze-in mechanism of DM genesis.

\section{Experimental constraints}
\label{sec:constraint}
The generation-specific FN charges in the Lagrangian (in Eq.~\eqref{eq:Lag}) induce flavor-changing neutral currents at the tree level via the exchange of (pseudo)scalar bosons. This effect is larger for heavier fermions, as can be seen from the interaction strengths quoted in Eqs.~\eqref{eq:g+} and \eqref{eq:g-}, which are proportional to the fermion masses. In this section, we explore the constraints on the model parameter space that arise from low- and high-energy flavor-changing processes. 

\subsection{Constraints from low-energy processes}

In the following subsections, we study the constraints on the model parameters obtained from the processes occurring at the energy scale below the mass of the pseudoscalar flavon.

\subsubsection{Meson mixing}
\label{sec:Meson}

We write the effective Hamiltonian relevant for the $B_{(s)}$-meson mixing involving the pseudoscalar $a$ (heavier than $m_b$) as follows.
\begin{align}
H_\text{eff}^{\Delta F=2} &= \frac{G_F^2}{16 \pi^2} M_W^2 (V_{tb}^* V_{tq})^2  C_1^{VLL}(\mu)\, Q_1^{VLL}(\mu) \nn \\
&+ C_1^{SLL}(\mu) \,Q_1^{SLL}(\mu)+ C_1^{VRR}(\mu)\, Q_1^{VRR}(\mu)+ C_2(\mu) \,Q_2(\mu) \,.
\end{align}
Within the SM only one single operator 
\begin{align}
Q_1^{VLL} \ = \ \left(\bar{b}^\alpha \gamma_\mu P_L q^\alpha  \right)\left(\bar{b}^\beta \gamma^\mu P_L q^\beta  \right)
\end{align}
is present, where $q=d,\,s$ for $B_d$-mixing and $B_s$-mixing, respectively.
The operators induced by the pseudoscalar interaction are given as
\begin{align}
Q_1^{SLL} \ &= \ \left(\bar{b}^\alpha P_L q^\alpha  \right)\left(\bar{b}^\beta  P_L q^\beta  \right)\,, \nn\\[1ex]
Q_1^{SRR} \ &= \ \left(\bar{b}^\alpha P_R q^\alpha  \right)\left(\bar{b}^\beta  P_R q^\beta  \right)\,, \nn\\[1ex]
Q_2^{LR} \ &= \ \left(\bar{b}^\alpha P_L q^\alpha  \right)\left(\bar{b}^\beta  P_R q^\beta  \right)\,.
\end{align}
The resultant contribution to the mass difference of the neutral meson can then be written as
\begin{align}
\label{eq:DeltaMq}
\Delta M_{q}\equiv \bigg|\frac{\langle \bar{B_q}|H_\text{eff}^{\Delta F=2}  |B_q \rangle }{M_{B_q}} \bigg| &= \bigg|1+ \frac{P_1^{SLL}\left(C_1^{SLL}+C_1^{SRR}  \right) + P_2^{LR} C_2^{LR} }{P_1^{VLL}\frac{G_F^2}{4 \pi^2} M_W^2 (V_{tb}^* V_{tq})^2 S_0(m_t^2/m_W^2)  }\bigg|\Delta M_{q}^\text{SM} \,, \\[2ex]
\text{where} \quad S_0(x) &= \dsp\frac{\frac{x^3}{4}-\frac{11 x^2}{4}+x}{(1-x)^2}-\frac{3 x^3 \log (x)}{2 (1-x)^3}\,,
\end{align}
is the Inami-Lim function denoting the SM contribution. 
The bag factors, introduced to parameterize the hadronic transition element as $\langle \bar{B_q}|Q_i |B_q \rangle = \dsp\frac{2}{3} M_{B_q}^2 f_{B_q}^2 P_i$, including the renormalization group evolution are estimated at $m_b$~\cite{Gorbahn:2009pp}:
\begin{align}
P_1^{VLL}=0.71\pm 0.05 ,\qquad P_1^{SLL} = -1.36\pm 0.12 ,\qquad P_2^{LR}=3.2\pm 0.2\,.
\end{align}
Note that $Q_1^{SLL}$ and $Q_1^{SRR}$ both have the same hadronic matrix element.
Now, the new physics Wilson coefficients arising from the pseudoscalar interaction are given by
\begin{align}
C_1^{SLL} \ &= \ -\frac{\left[(g_-^d)_{32}-(g_+^d)_{32}  \right]^2}{ m_a^2}\,,\nn\\[1ex]
C_1^{SRR} \ &= \ -\frac{\left[(g_-^d)_{32}+(g_+^d)_{32}  \right]^2}{ m_a^2}\,,\nn\\[1ex]
C_2^{LR} \ &= \ \frac{\left[(g_-^d)_{32}+(g_+^d)_{32}  \right]\left[(g_-^d)_{32}-(g_+^d)_{32}  \right]^*}{ m_a^2} \,.
\end{align}
The current average of the theoretical predictions compared to the measurements is found to be~\cite{DiLuzio:2019jyq}
\begin{align}
\Delta M_d^\text{average} \ &= \ \left(1.05^{+0.04}_{-0.07}\right) \Delta M_d^\text{exp}\,, \nn\\[1ex]
\Delta M_s^\text{average} \ &=\ \left(1.04^{+0.04}_{-0.07}\right) \Delta M_s^\text{exp} \,,
\end{align}
which, by Eq.~\eqref{eq:DeltaMq}, imposes 
\begin{align}
v_\phi \, m_a \ \gtrsim \ 1.8 \times 10^5 \, \text{GeV}^2,
\label{eq:B-mixing}
\end{align}
with $m_a>m_b$. This bound significantly limits the FN symmetry breaking scale $v_\phi$, especially for $m_a \gtrsim \mathcal{O}(10)$ GeV where the constraints from meson decays are irrelevant. Note that the scalar flavon $s$ also contributes to these processes. However, as $m_s = 2 \sqrt{\lambda_\phi} \, v_\phi$, these effects are highly suppressed (with a factor of $\sim1/v_\phi^2$) compared to the bound mentioned (in Eq.~\eqref{eq:B-mixing}) arising from the pseudoscalar flavon. We find that the bounds arising from the light-meson mixing data, such as kaon and $D^0$-meson, are less stringent than the one quoted above for $B_s$-mixing.

\subsubsection{Radiative lepton decay}
\label{sec:muegamma}

\begin{figure}[H]
\begin{center}
\includegraphics[trim={0cm 0cm 0cm 0cm },clip, width=0.8\textwidth]{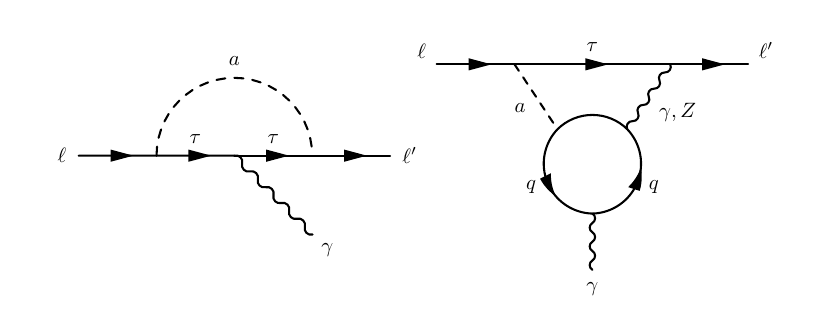}
\end{center}
\vspace{-0.5cm}
\caption{The Feynman diagrams for $\ell \to \ell^\prime \gamma$: (left) one-loop diagram and (right) two-loop Barr-Zee diagram.}
\label{fig:mutoe}
\end{figure}

Radiative lepton-flavor violating decays $\ell \to \ell^\prime \gamma$ are highly suppressed in the SM and stringent bounds on the decays, especially with light leptons, exist from the experiments. Although the flavon does not directly couple to the SM gauge bosons at tree level, the decays $\ell \to \ell^\prime \gamma$ can be generated at one- and two-loop levels, as shown in Fig.~\ref{fig:mutoe}.

The decay rate of $\ell \to \ell^\prime \gamma$ can be written as~\cite{Harnik:2012pb}
\begin{equation}
\Gamma(\ell \rightarrow \ell^\prime \gamma)=\frac{\alpha m_\ell^5}{64 \pi^4}\left(\left|\mathcal{A}_L\right|^2+\left|\mathcal{A}_R\right|^2\right)\,,
\end{equation}
where the amplitudes $\mathcal{A}_{L,R}=\mathcal{A}_{L,R}^{1\,{\rm loop}}+ \mathcal{A}_{L,R}^{2\,{\rm loop}} $ include both one- and two-loop contributions, respectively.
In the case of $\mu \to e \gamma$, at one-loop level the tau contribution is dominant which in the limit $m_\mu \ll m_\tau \ll m_a$ can simply be written as
\begin{align}
\mathcal{A}_{L}^{1\,{\rm loop}}\big|_{\mu \to e \gamma} \simeq \frac{1}{8m_a^2} \frac{m_\tau}{m_\mu} (g^{e})_{21}^*(g^e)_{33}  \left[ -3 + 2 \log \left( \frac{m_a^2}{m_\tau^2} \right)  \right]\,, \\
\mathcal{A}_{R}^{1\,{\rm loop}}\big|_{\mu \to e \gamma} \simeq \frac{1}{8m_a^2} \frac{m_\tau}{m_\mu} (g^e)_{12}(g^e)_{33}   \left[ -3 + 2 \log \left( \frac{m_a^2}{m_\tau^2} \right)  \right]\,,
\end{align}
where $(g^f)_{ij}\equiv (g_+^f)_{ij}+(g_-^f)_{ij},~ (g^f)^*_{ji} \equiv (g_+^f)_{ij}-(g_-^f)_{ij}$ can be derived from Eqs.~\eqref{eq:g+} and \eqref{eq:g-}. With appropriate replacements of coupling constants and expansion of the loop functions, the one-loop amplitudes for $\tau \to \mu \gamma$ which is also dominated by the tau loop, read
\begin{align}
\mathcal{A}_{L}^{1\,{\rm loop}}\big|_{\tau \to \mu \gamma} \simeq \frac{1}{12m_a^2}  (g^{e})_{32}^*(g^e)_{33}   \left[ -4 + 3 \log \left( \frac{m_a^2}{m_\tau^2} \right)  \right]\,, \\
\mathcal{A}_{R}^{1\,{\rm loop}}\big|_{\tau \to \mu \gamma} \simeq \frac{1}{12m_a^2}  (g^e)_{23}(g^e)_{33} \left[ -4 + 3 \log \left( \frac{m_a^2}{m_\tau^2} \right)  \right]\,.
\end{align}

We then include the two-loop contribution arising from the Barr-Zee diagram, where all six flavors of the quarks contribute while the bottom and top quarks are the dominant ones.\footnote{Note that, despite the vanishing FN charge for the top quark, the flavon coupling to the top quark is generated in the mass basis after performing the rotation. This has sometimes been overlooked in the literature when considering constraints from low-energy processes~\cite{Bauer:2016rxs}.}

\begin{align}
\label{eq:2loop}
\mathcal{A}_{L}^{2\,{\rm loop}}\big|_{\ell \to \ell^\prime \gamma} = \frac{6\alpha_{\rm EM}}{\sqrt{2}\pi} \frac{G_F^2 v_{\rm EW}^2}{m_\ell} \sum_{\substack{f=u,d\\ i=1,2,3}} \frac{Q_i^{f2}}{m_i^f} 
 (g^{e})_{\ell\ell^\prime}^*(g^f)_{ii} \, \mathcal{G}\left( \frac{m_i^{f2}}{m_a^2} \right)\,,
\end{align}
where the loop function is given by
\begin{equation}
\label{eq:Gloop}
\mathcal{G}(z)=\frac{1}{2} z \int_0^1 d x \frac{1}{x(1-x)-z} \log \frac{x(1-x)}{z}\,.
\end{equation}
The expression for $\mathcal{A}_{R}^{2\,{\rm loop}}\big|_{\ell \to \ell^\prime \gamma}$ can be obtained by replacing $(g^{e})_{\ell\ell^\prime}^*$ in Eq.~\eqref{eq:2loop} with $(g^{e})_{\ell^\prime\ell}$.

The MEG experiment provides the most stringent upper limit on $\mu \to e \gamma$ \cite{MEG:2016leq}, while for $\tau \to \ell \gamma$ the strongest bounds have been placed by the BaBar collaboration \cite{BaBar:2009hkt}. The current 90\% C.L. limits are:
\begin{eqnarray}
{\rm BR}(\mu \to e \gamma) < 4.2\times 10^{-13},~~ {\rm BR}(\tau \to e \gamma)<3.3 \times 10^{-8},~~ {\rm BR}(\tau \to \mu \gamma)<4.4 \times 10^{-8}\,.
\end{eqnarray}

The parametric form of the one-loop contribution to the $\mu \to e \gamma$ branching fraction in our model can be written as
\begin{equation}
{\rm BR}^{1\,{\rm loop}}\big|_{\mu \to e \gamma}   \simeq 2.4 \times 10^6\, \frac{\left[ -3 + 2 \log (0.3\, m_a^2)\right]^2}{m_a^4 v_\phi^4}\,,
\end{equation}
which turns out to be one order of magnitude larger than the two-loop Barr-Zee contribution as long as $m_a<200\,\gev$. However, these two contributions are comparable for larger $m_a$, and we include both of these contributions in our analysis. The reason being that the loop function $\mathcal{G}(z)$ (in Eq.~\eqref{eq:2loop}) decreases slowly with $m_a$ compared to the $1/m_a^2$ drop-off of the one-loop amplitude.
We will see in the upcoming section~\ref{sec:freeze-out} that for the choice of our parameter space, $\mu \to e \gamma$ remains the most stringent constraint for the region of low flavon mass ($\sim\mathcal{O}(100)\,\gev$) with $v_\phi\sim \mathcal{O}(1)\,\tev$. The future sensitivity of MEG-II is expected to be ${\rm BR}(\mu \to e \gamma) \sim 6 \times 10^{-14}$~\cite{MEGII:2018kmf} which will be effective in stretching the bounds further in the TeV range.

\subsection{Constraints from high energy processes: top quark decay}
\label{sec:Top}

Depending on the mass scale of the pseudoscalar, it can decay to different sets of fermions, which will lead to possible signatures at the colliders. 
The decay width of the pseudoscalar to fermions i.e., $a \to \bar{f}^i f^j$ is given by
\begin{align}
    \Gamma_a^{ij} = & \ \frac{N_c}{8 \pi m_a^3 } \lambda^{1/2} (m_a^2,m_{f_i}^2,m_{f_j}^2) \ \times \nn\\[1ex]
    &\left[ |(g_-^f)_{ij}|^2 \left( 1 - \dsp\frac{(m_{f_i}+m_{f_j})^2}{4m_a^2} \right)+|(g_+^f)_{ij}|^2 \left( 1 - \dsp\frac{(m_{f_i}-m_{f_j})^2}{4m_a^2} \right) \right] \,,
\end{align}
where $\lambda(x,y,z)=x^2+y^2+z^2 - 2xy-2yz-2zx$ is the Kall\'en function and the color factor $N_c=3(1)$ for the quarks(leptons).
\begin{figure}[h]
	\begin{center}
		\includegraphics[trim={0cm 0cm 0cm 0cm },clip, width=0.57\textwidth]{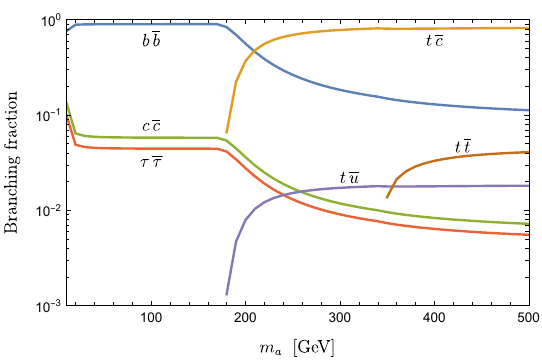}
	\end{center}
	\vspace{-0.5cm}
	\caption{
		The branching fractions of the pseudoscalar flavon decay into the SM fermions for the mass range $m_a\in\,[10-500]\,$GeV. Only those decay channels with branching fractions greater than 1\% are shown in the plot. For different flavor final states, the charge-conjugate mode is also included. DM $N^1$ is assumed to be heavy enough so that the $a \to N^1 N^1$ process is kinematically forbidden.
	}\label{fig:ALP-BR}
\end{figure}

We show the branching fractions to dominant channels involving SM fermions in Fig.~\ref{fig:ALP-BR} for the pseudoscalar in the mass range $10\,\gev$ to $500\,$GeV. For different flavor final states, the charge-conjugate mode is also included. We notice that before the opening of the top-quark pair threshold, the decay of $a\to b\bar{b}$ almost saturates the total rate. This feature is expected, since the coupling of the pseudoscalar to the fermions is proportional to the masses of the fermions. Among the leptons, only the $a\to \tau \bar{\tau}$ channel is notable, since $\tau$ is the heaviest lepton. We have assumed that the DM is heavy enough so that the $a \to N^1 N^1$ process is kinematically forbidden.

We infer from the above discussion that in the model of our consideration, if the pseudoscalar is lighter than the top quark, the top quark can decay into the pseudoscalar and SM fermions at tree level, and a subsequent decay of the pseudoscalar to the $b \bar{b}$ pair can be a very promising signature at colliders due to high branching rates of these two processes. 
This decay is indeed sought and constrained by a recent search at the LHC. The ATLAS collaboration~\cite{ATLAS:2023mcc} presents a general search for the production of top quark pairs with the full Run 2 dataset at 13 TeV center-of-mass energy, where one of the top quarks decays into either an up quark or a charm quark, and a light scalar particle $X$, with $X \to b \bar{b}$ subsequently. The other top quark decays to a $W$-boson and a $b$-quark according to the SM.

\begin{figure}[ht]
	\begin{center}
		\includegraphics[trim={0cm 0cm 0cm 0cm},clip, width=0.6\textwidth]{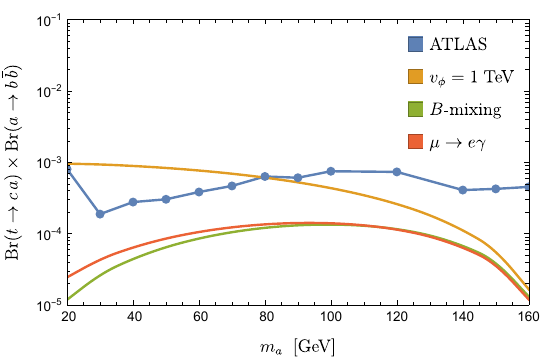}
	\end{center}
	\vspace{-0.5cm}
	\caption{
		Constraints on the product ${\rm BR}(t \to c a) \times {\rm BR}(a \to b \bar{b})$ for $m_a\in\,[20-160]\,$GeV. The blue dots represent the upper limits at 95\% C.L. from the ATLAS search~\cite{ATLAS:2023mcc}. Our model predicts the yellow curve for a benchmark choice $v_\phi = 1\,$ TeV which is excluded by ATLAS data for $m_a\lesssim 80\,\gev$. The green and red curves are the indirect upper bounds on this product of branching fractions arising from $B$-meson mixing and $\mu \to e \gamma$, respectively, which turn out to be more constraining than direct searches (see the text for details).
	}\label{fig:ATLAS}
\end{figure}

In our setup, the charm quark interacts with the pseudoscalar and top quark with a coupling much larger than that of the up quark due to the relatively heavy mass of the charm. Therefore, the ATLAS search imposes a stronger limit on the product ${\rm BR}(t \to c a) \times {\rm BR}(a \to b \bar{b})$ predicted in our model, which in turn translates into the constraint on the FN symmetry breaking scale $v_\phi$. We present these constraints along with the bounds from $B$-meson mixing and $\mu \to e \gamma$ in Fig.~\ref{fig:ATLAS}, for the pseudoscalar mass ranging from 20\,GeV to 160 GeV.
The blue dots in Fig.~\ref{fig:ATLAS} denote the upper limits at 95\% C.L. on the product of the decay branching fractions ${\rm BR}(t \to c X) \times {\rm BR}(X \to b \bar{b})$ presented in the aforementioned ATLAS search, where $X$ is a light scalar particle. The yellow curve denotes the product ${\rm BR}(t \to c a) \times {\rm BR}(a \to b \bar{b})$ for the choice $v_\phi = 1\,$TeV in our model. The green and red curves illustrate this product of branching fractions after including the constraints from the $B$-meson mixing obtained in Sec.~\ref{sec:Meson} and $\mu \to e \gamma$ obtained in Sec.~\ref{sec:muegamma}. To elaborate, for each $m_a$, the lower bound on $v_\phi$ from $B$-meson mixing can be obtained with Eq.~\eqref{eq:B-mixing}, that is, $ v_\phi > 1.8 \times 10^5/m_a \, \text{GeV}^2$, which leads to an upper bound on the product ${\rm BR}(t \to c a) \times {\rm BR}(a \to b \bar{b})$. The bounds from $\mu \to e \gamma$ are derived in a similar manner.
With Fig.~\ref{fig:ATLAS} we infer that the constraint on $v_\phi$ from the top quark decay at the LHC is not as stringent as the ones from the $B$-meson mixing or $\mu \to e \gamma$. For $v_\phi = 1\,$TeV, $m_a > 80\,$GeV is still allowed by the ATLAS search. However, it is ruled out by both the $B$-meson mixing data and the experimental upper limit of $\mu \to e \gamma$.

\section{Dark Matter genesis and relic density}

The thermal history, interaction strength, and the mass range of the DM particle govern the mechanisms that give rise to the relic density of the Universe. One of the most popular mechanisms is the Weakly Interacting Massive Particle (WIMP) scenario, where the mass of the DM lies within the range $\mathcal{O}(1)\,$GeV -- $\mathcal{O}(10)\,$TeV and interacts with the SM particles via a weak interaction~\cite{Hut:1977zn,Lee:1977ua,Vysotsky:1977pe,Srednicki:1988ce,Gondolo:1990dk}. The WIMP DM is produced via thermal freeze-out and has many implications for a diverse group of search experiments, such as direct and indirect detection and colliders. An alternative mechanism, based on the thermal freeze-in scenario~\cite{McDonald:2001vt,Hall:2009bx,Bernal:2017kxu}, suggests that the DM particle was never in equilibrium with the thermal bath and was produced from a mediator, which in our case can be flavon fields. Due to its feeble interaction nature, it can escape traditional detections, mainly designed for WIMP setups. In the following, we explore both of these two possibilities of DM genesis in the context of the model considered in Sec.~\ref{sec:model}.

\subsection{Freeze-out scenario}
\label{sec:freeze-out}

\begin{figure}[H]
\begin{center}
\includegraphics[trim={0cm 0cm 0cm 0cm },clip, width=0.35\textwidth]{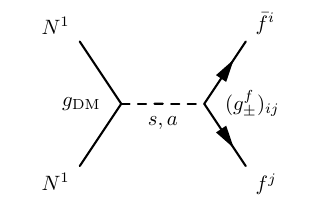}
\end{center}
\vspace{-0.5cm}
\caption{The $s$-channel Feynman diagram for DM annihilation process $N^1 N^1 \to \bar{f}^i f^j$.}
\label{fig:NN-ff}
\end{figure}

If the FN symmetry breaking scale $v_\phi$ is around the weak scale, say $v_\phi \sim \mathcal{O}(\text{TeV})$, the RHN DM $N^1$ may annihilate to the SM fermions and flavon particles. As the Universe expands and cools down, their number density decreases. They may eventually freeze out and contribute to the observed DM relic density. $N^1$ interacts with SM fermions through the scalar and pseudoscalar portal through the $s$-channel processes, as shown in Fig.~\ref{fig:NN-ff}.

As shown in Eq.~\eqref{eq:masses}, the scalar $s$ is much heavier than the pseudoscalar $a$, the contribution of the scalar to the above annihilation process will only be effective in the high mass range of $N^1$. From Eq.~\eqref{eq:gaf} and Eq.~\eqref{eq:gDM} we can see that the coupling strength between the scalar and pseudoscalar with the fermions and RHNs are the same up to an overall phase factor and are proportional to the fermion and RHN masses, respectively. To have efficient annihilation processes, RHN $N^1$ is supposed to be heavier than SM fermions, that is, $|q_{N_1}| \times \mdm \gtrsim 100\,$ GeV, which implies that $\gdm > \gff$. The cross section of the $N^1 N^1 \to \bar{f}^i f^j$ process is proportional to the square of both couplings
$\sigma_{N^1N^1 \to f^if^j} \ \sim \ \gdm^2 \times \gff^2\,$.

\begin{figure}[H]
\begin{center}
\includegraphics[trim={0cm 0cm 0cm 0cm },clip, width=0.85\textwidth]{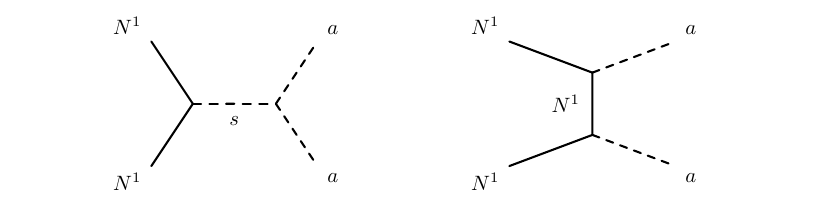}
\end{center}
\vspace{-0.5cm}
\caption{The $s$-channel (left panel) and $t$-channel (right panel) Feynman diagrams for DM annihilation process $N^1 N^1 \to a \, a$.}
\label{fig:NN-aa}
\end{figure}

In addition to the SM final states, the DM $N^1$ can also annihilate to the flavon particles, and the latter will consecutively decay into SM fermions and hence facilitate the freezing out of the DM. For example, the process $N^1 N^1 \to a \, a$ can take place via the $s$-channel and the $t$-channel, as shown in the diagrams in Fig.~\ref{fig:NN-aa}.
The contribution to the total cross section of $N^1 N^1 \to a \, a$ from the $t$-channel is parametrically larger than that of the $N^1 N^1 \to \bar{f^i} f^j$ process due to the fact that $\gdm > \gff$ and the cross section in the prior case varies as
$\sigma_{NN \to aa}^{t-channel} \ \sim \ \gdm^4 \ > \ \gff^2 \times \gdm^2 \ \sim \ \sigma_{N^1N^1 \to f^if^j}
$. In addition to the two pseudoscalars as the final state, $N^1$ can also be annihilated into one scalar and one pseudoscalar ($N^1 N^1 \to s \, a$) or two scalars ($N^1 N^1 \to s \, s$) if kinematically allowed. 

We calculate the amplitudes and cross sections for all possible $N^1$ annihilation processes in the model under consideration.
The expressions are provided in Appendix~\ref{app:Xsec} and also verified by implementing the model in \texttt{FeynRules} \cite{Christensen:2008py,Alloul:2013bka} with the help of \texttt{FeynArts} \cite{Kublbeck:1990xc,Hahn:2000kx} and \texttt{FeynCalc} \cite{Mertig:1990an,Shtabovenko:2020gxv}.

The thermally averaged cross section for $N^1$ annihilation is given in~\cite{Gondolo:1990dk}:
\begin{align}
\label{eq:sigmav}
\sv \ = \ \frac{1}{8\mdm^4 T K_2^2 (\mdm / T)} \int_{4\mdm^2}^{\infty} d\ecm^2 \ \sigma(\ecm) \left(\ecm^2 - 4\mdm^2\right) \ecm \, K_1\left(\frac{\ecm}{T}\right) \,,
\end{align}
where $T$ is the freeze-out temperature, $E_\text{CM}$ is the center-of-mass energy, $\sigma(E_\text{CM})$ is the total cross section of the annihilation process at $E_\text{CM}$, $K_1$ and $K_2$ are the modified Bessel functions of order one and two, respectively. The freeze-out temperature $T$ is calculated numerically. Finally, the relic abundance of DM is calculated by solving the Boltzmann equation with the thermally averaged cross section $\sv$ obtained in Eq.~\eqref{eq:sigmav}.

\begin{figure}[ht]
	\begin{center}
		\includegraphics[trim={0 0 0 0},clip, width=0.6\textwidth]{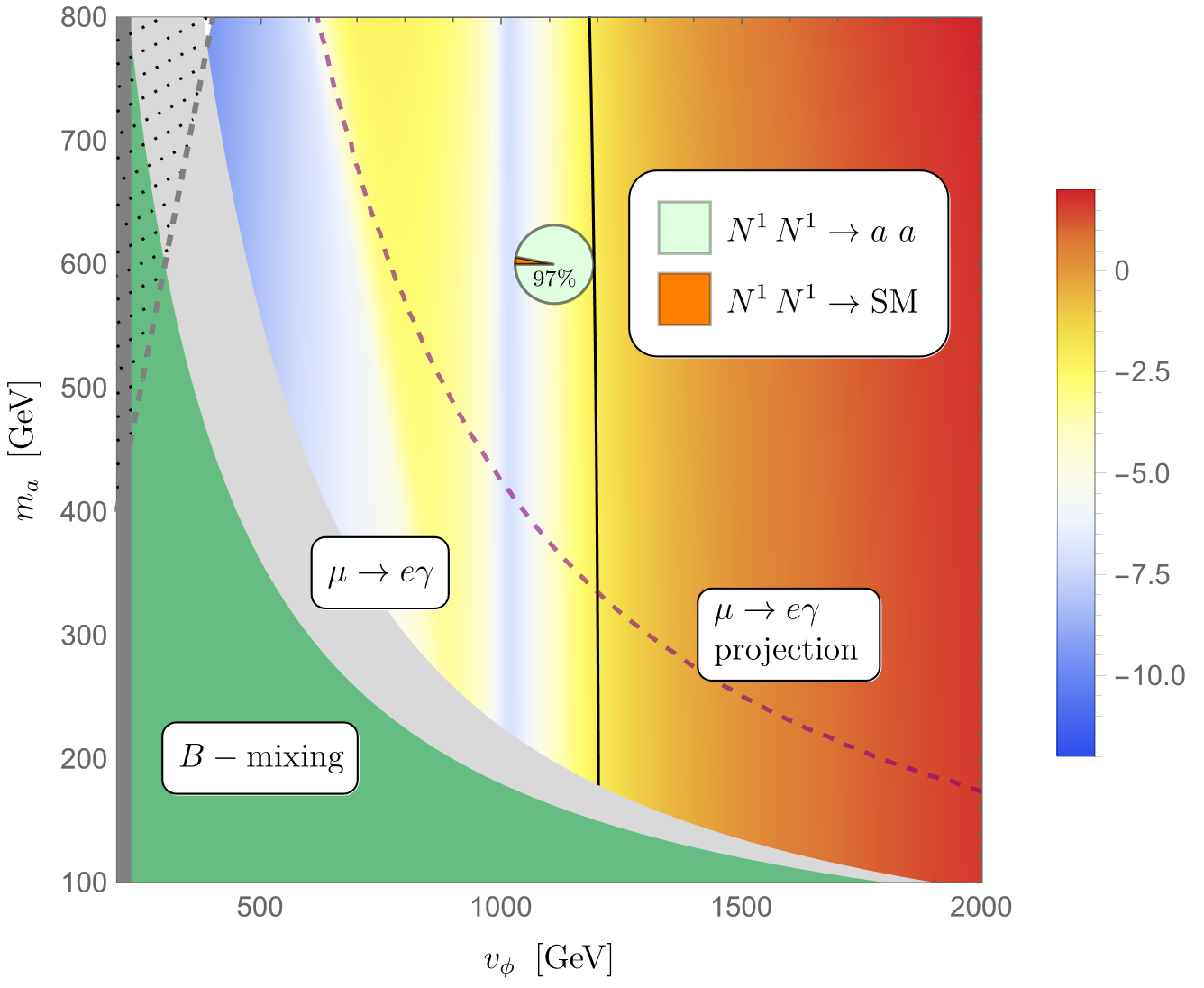}
	\end{center}
	\vspace{-0.5cm}
	\caption{
		Color plot for the logarithm of the relic density, i.e. $\ln \omg$ for $m_a \in \, [100-800]\,$ GeV and $v_\phi \in \, [200-2000]\,$ GeV, with $\mdm = 1\,$ TeV and $\lambda_\phi = 1$. The black line indicates the observed DM relic abundance $\omg = 0.120 \pm 0.001$~\cite{Planck:2018vyg}, which is a result dominated by the freeze-out process $N^1 N^1 \to a \,a$. The green region is excluded by the $B$-meson mixing data (Eq.~\eqref{eq:B-mixing}). The curved gray region is excluded by the limits of the branching fraction $\mu \to e \gamma$ of the MEG experiment~\cite{MEG:2016leq} and the future sensitivity of MEG-II~\cite{MEGII:2018kmf} is shown in dashed purple. The dark-gray region of $v_\phi < 230\,$GeV is excluded by the validity of the effective theory. The triangle region to the left of the dashed gray is not viable because $m_a > m_s$. As illustrated by the pie chart, the contribution of $N^1 N^1 \to a \,a$ is 97\%.
	}\label{fig:vphi-ma}
\end{figure}

For the estimation of the relic abundance, we implement our model in \texttt{MicrOMEGAs} \cite{Belanger:2006is,Belanger:2013oya,Belanger:2018ccd} and \texttt{MadDM} \cite{Ambrogi:2018jqj,Arina:2021gfn}, and scan various parameter spaces.\footnote{As a cross-check, we compared the outputs from the two different packages with our own calculation. The results are consistent with each other.}
The relic abundance of DM is governed by three parameters in our model, namely the mass of DM $\mdm$, the pseudoscalar mass $m_a$, and the FN symmetry breaking scale $v_\phi$. The scalar mass $m_s = 2 \sqrt{\lambda_\phi} \, v_\phi$, as shown in Eq.~\eqref{eq:masses}. First, we choose $\lambda_\phi = 1$ as our benchmark. Later (in Fig.~\ref{fig:diff-lam}) we will repeat our analysis for five reference cases of $\lambda_\phi = \{ 0.1,\ 0.3,\ 0.5,\ 2,\ 3 \}$. To investigate viable DM freeze-out scenarios that meet the current phenomenological constraints discussed in Sec.~\ref{sec:constraint}, we choose the following range for the mass spectrum of the particles and the FN symmetry breaking scale. 
\begin{align}
m_a  \sim \mathcal{O}(100)\,\gev, \qquad \mdm,~v_\phi \sim  \mathcal{O}(1-10)\,\tev.
\end{align}

Our results indicate that the relic abundance is not sensitive to the mass of the pseudoscalar $m_a$, as long as $m_a$ stays below the mass of $N_1$. In Fig.~\ref{fig:vphi-ma}, we show a scan over the parameter space for $\lambda_\phi = 1$ and $N_1$ at 1 TeV, with the pseudoscalar mass $m_a$ varying from 100\,GeV to 800\,GeV and the FN symmetry breaking scale $v_\phi$ from 200\,GeV to 2\,TeV. We denote the logarithm of the relic density $\ln \omg$ by the color, as shown in the legend.

The black line indicates the observed DM relic abundance from the PLANCK experiment~\cite{Planck:2018vyg}, $\omg = 0.120 \pm 0.001$, achieved at $v_\phi = 1.2\,$TeV. The contribution is dominated (97\%) by the $N_1$ annihilation process to pseudoscalars $N^1 N^1 \to a \,a$, as illustrated by the pie chart in Fig.~\ref{fig:vphi-ma}. The annihilation to the SM final states (predominantly to the $t \, \bar{c}$ and $\bar{t} \, c$ states) contributes only at the 3\% level. The curved green region of $v_\phi m_a < 1.8 \times 10^5 \, \text{GeV}^2$ has been excluded by the $B$-meson mixing data, as mentioned in Section~\ref{sec:Meson}. The curved gray region is excluded by the limits of the branching fraction $\mu \to e \gamma$ of the MEG experiment~\cite{MEG:2016leq} and the future sensitivity of MEG-II~\cite{MEGII:2018kmf} is shown in dashed purple. The dark-gray region of $v_\phi < 230\,$GeV is excluded by the validity of the effective theory. For $\mdm = 1\,$TeV, we require that
\begin{align}
\frac{v_\phi}{\epsilon} = M > \mdm \quad \Longrightarrow \quad v_\phi > (\epsilon \times \mdm) = 0.23\, \mdm = 230 \, \text{GeV}\,.
\label{eq:EFT}
\end{align}
Furthermore, the triangle region to the left of the dashed gray is not viable because $m_a > m_s$. Note that the relic abundance is not sensitive to the pseudoscalar mass in this range.

\begin{figure}[ht]
	\begin{center}
		\includegraphics[trim={0cm 0 0 0},clip, width=0.6\textwidth]{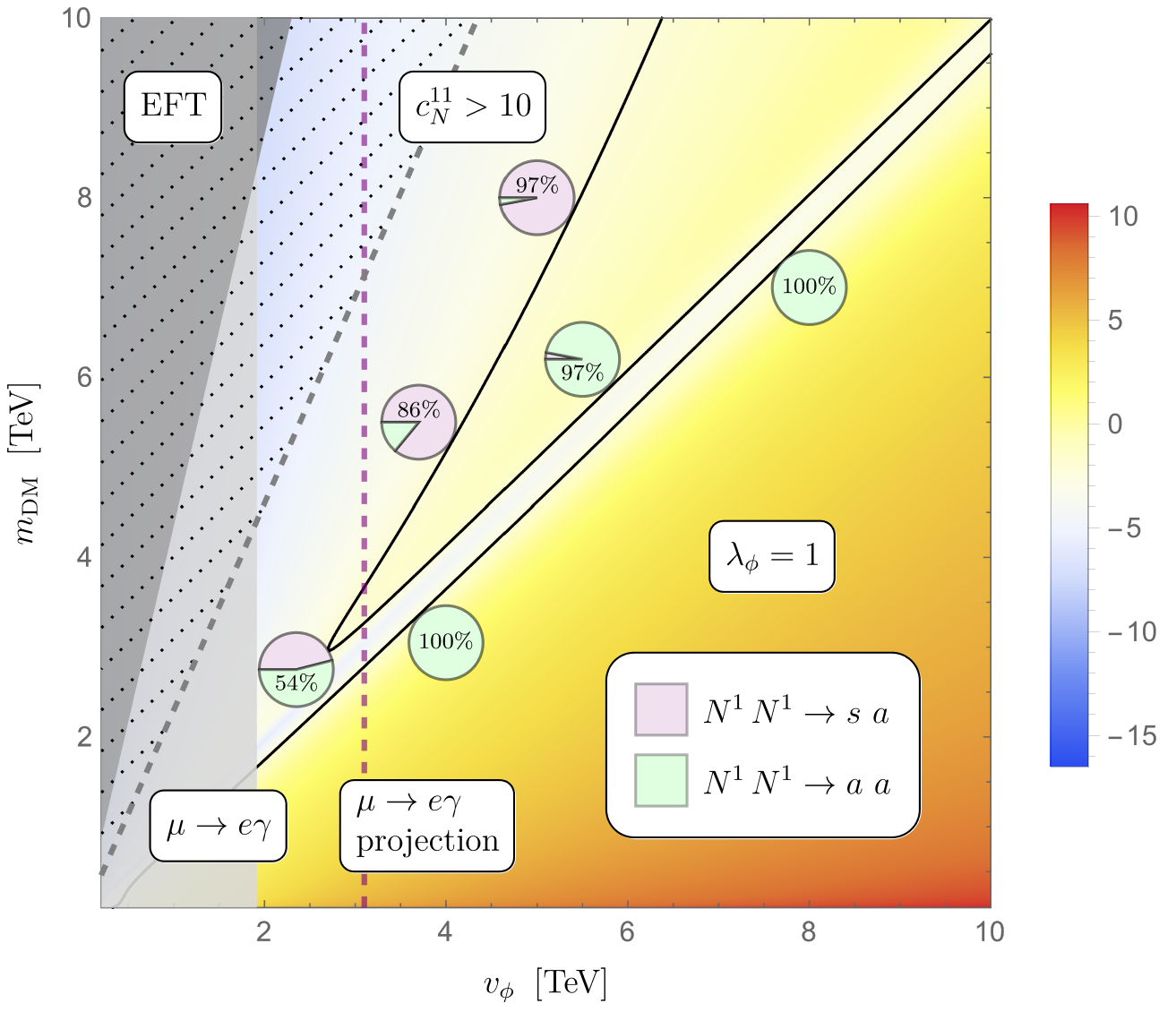}
	\end{center}
	\vspace{-0.5cm}
	\caption{
	Contour plot for the logarithm of relic density, i.e. $\ln \omg$ for $\mdm \in\,[0.1-10]\,$TeV and $v_\phi \in\,[0.2-10]\,$TeV, with $m_a = 100\,$GeV and $\lambda_\phi = 1$ as a benchmark case. Black contour lines indicate the observed DM relic abundance $\omg = 0.120 \pm 0.001$~\cite{Planck:2018vyg}. The light-gray region is excluded by the limits of the branching fraction $\mu \to e \gamma$ of the MEG experiment~\cite{MEG:2016leq} and the future sensitivity of MEG-II~\cite{MEGII:2018kmf} is shown in dashed purple. The dark-gray triangle is excluded by the validity of effective theory. The region to the left of the dashed gray is excluded because it indicates $c_N^{11} > 10$, which is disfavored by the FN framework. Pie charts denote the contribution from the dominant freeze-out channel in percentage, specifically at the point where the contour line intersects with the pie chart.
	}\label{fig:freeze-out-1}
\end{figure}

Next, we focus on exploring the range of the two relevant parameters that the relic abundance is sensitive to, namely the $N_1$ mass $\mdm$ (from 100 GeV to 10 TeV) and the FN symmetry-breaking scale $v_\phi$ (from 200 GeV to 10 TeV). We fix the pseudoscalar mass $m_a$ at 100 GeV to reduce the impact of the low-energy constraints.

In Fig.~\ref{fig:freeze-out-1}, we plot the logarithm of the relic density of DM, $\ln \omg$, where the black contour lines indicate the observed value of DM relic abundance. The light-gray region is excluded by the limits of the branching fraction $\mu \to e \gamma$ of the MEG experiment~\cite{MEG:2016leq} and the future sensitivity of MEG-II~\cite{MEGII:2018kmf} is shown in dashed purple. The bound derived from the mixing of $B$-mesons quoted in Eq.~\eqref{eq:B-mixing} is slightly less constraining than the current limit of $\mu \to e \gamma$ as demonstrated in Fig.~\ref{fig:freeze-out-1} and therefore we do not show it here. The dark-gray triangle is excluded by not satisfying the validity of effective theory (Eq.~\eqref{eq:EFT}), that is, $\mdm < M = v_\phi / 0.23$. The region to the left of the dashed gray is excluded because it indicates $c_N^{11} > 10$, a scenario disfavored within the FN framework. This framework adheres to the principle, from a naturalness point of view, of $\mathcal{O}(1)$ couplings at the Lagrangian level, with any modifications attainable by power counting in the parameter $\epsilon$.

There are three segments of black contours in Fig.~\ref{fig:freeze-out-1} where the observed relic abundance is obtained. Two annihilation processes, $N^1 N^1 \to s \,a$ and $N^1 N^1 \to a \,a$, dominate the contribution to the relic abundance in different regions of the parameter space. We draw pie charts to illustrate the contribution from the dominant freeze-out channel in percentage, specifically at the point where the contour line intersects with the pie chart.

The process $N^1 N^1 \to a \,a$ dominates the diagonal region of the parameter space that contains two parallel segments of black contours, which we call the "\textit{aa-branch}". Near the diagonal region, where $\mdm \approx v_\phi$, the process of the $s$-channel $N^1 N^1 \to a \,a$ is resonantly enhanced because the scalar, as mediator of this channel, has mass
\begin{align}
    m_s = 2 \sqrt{\lambda_\phi} \, v_\phi = 2 \, v_\phi \approx 2 \, \mdm \approx \ecm
\end{align}
for $\lambda_\phi = 1$, see the left panel of Fig.~\ref{fig:NN-aa} for the corresponding Feynman diagram. Therefore, the cross section of $N^1 N^1 \to a \,a$ is sufficiently large near the resonance to achieve the observed relic abundance of DM, hence dominating the \textit{aa-branch}.

The other process, $N^1 N^1 \to s \, a$, dominates the upper region of the parameter space that contains the other segment of the black contour, which we call the "\textit{sa-branch}". In this region, $\mdm > v_\phi$, so that $m_s < 2 \, \mdm \approx \ecm$ allows the scalar to be produced without kinematic suppression. As $\mdm$ and $v_\phi$ decrease, the contribution of $N^1 N^1 \to s \,a$ also decreases gradually along the \textit{sa-branch}, while that of $N^1 N^1 \to a \,a$ increases. At $\mdm = 3\,$TeV and $v_\phi = 2.7\,$TeV, the \textit{sa-branch} and the \textit{aa-branch} merge together. Contributions from $N^1 N^1 \to s \,a$ (46\%) and $N^1 N^1 \to a \,a$ (54\%) are roughly equal at this point. After merging, the total cross section receives contributions from both processes, resulting in the DM being overly annihilated and causing an underabundance of the relic density.

We infer from Fig.~\ref{fig:freeze-out-1}  that the DM freeze-out process is dominated by the contributions of annihilation of $N^1$ to flavon particles instead of SM fermions. Processes with different flavon final states dominate different regions of the parameter space, depending on the details of the model. It is worth noting that the relation $m_s = 2\,v_\phi$ is an artifact of our benchmark choice $\lambda_\phi = 1$. In Fig.~\ref{fig:diff-lam}, we show the scans on the same parameter space with six different choices of $\lambda_\phi = \{ 0.1,\ 0.3,\ 0.5,\ 1,\ 2,\ 3 \}$. The results qualitatively resemble those of the case $\lambda_\phi = 1$, although the details are different.

\begin{figure}[htbp]
    \centering
    \begin{subfigure}{0.4\textwidth}
        \centering
        \includegraphics[trim={0cm 0cm 0cm 0cm}, clip, width=\linewidth]{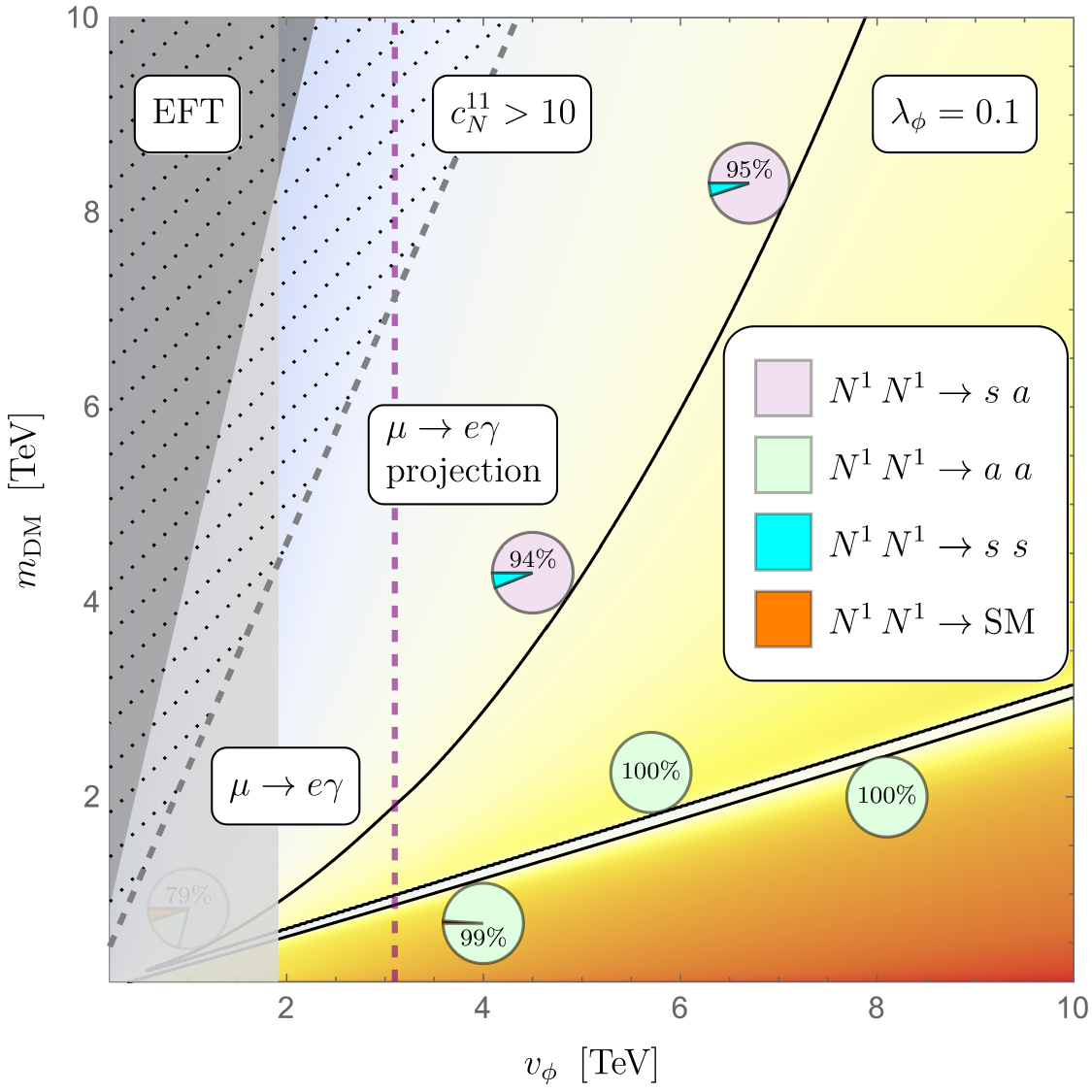}
    \end{subfigure}
    \begin{subfigure}{0.4\textwidth}
        \centering
        \includegraphics[trim={0cm 0cm 0cm 0cm}, clip, width=\linewidth]{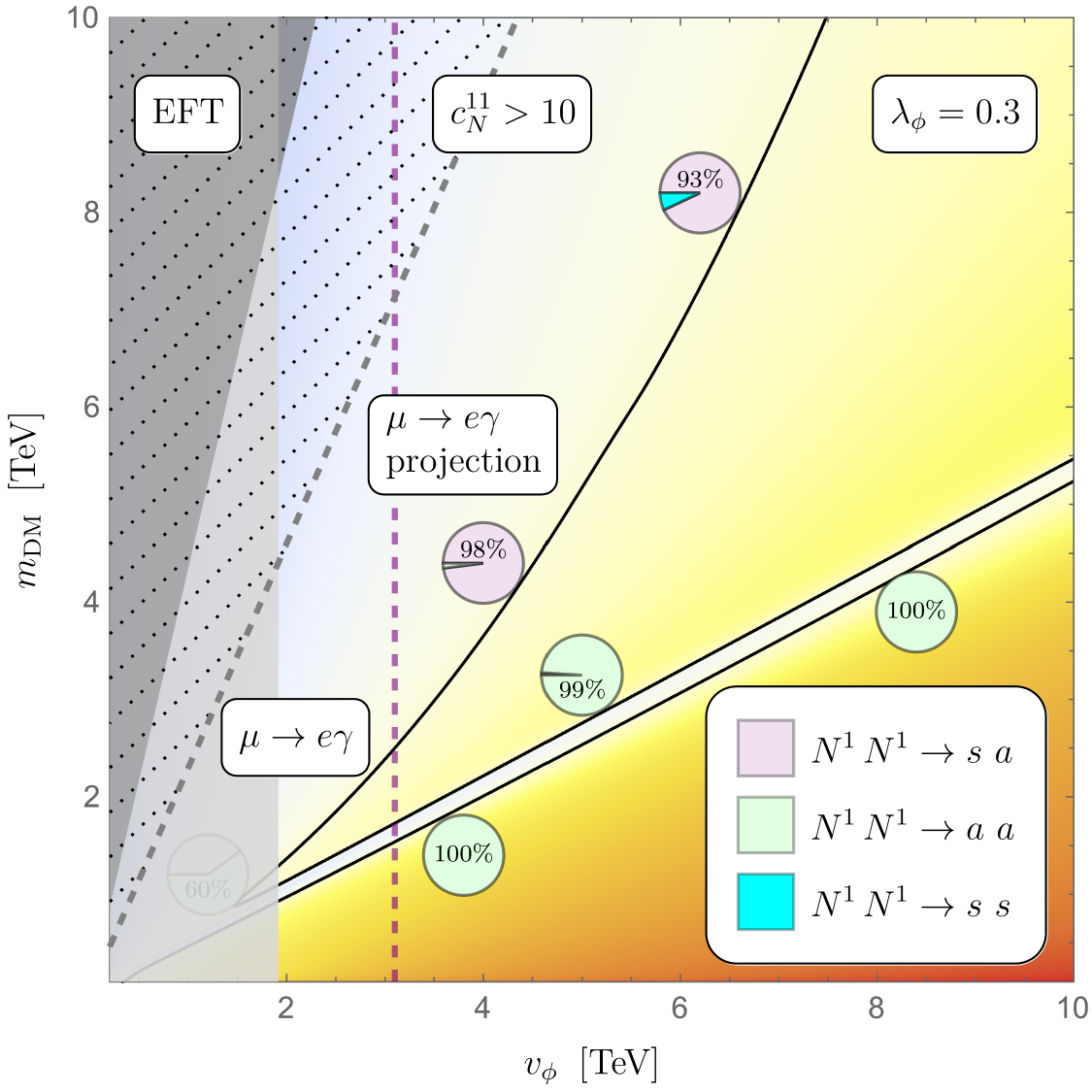}
    \end{subfigure}
    \\
    \begin{subfigure}{0.4\textwidth}
        \centering
        \includegraphics[trim={0cm 0cm 0cm 0cm}, clip, width=\linewidth]{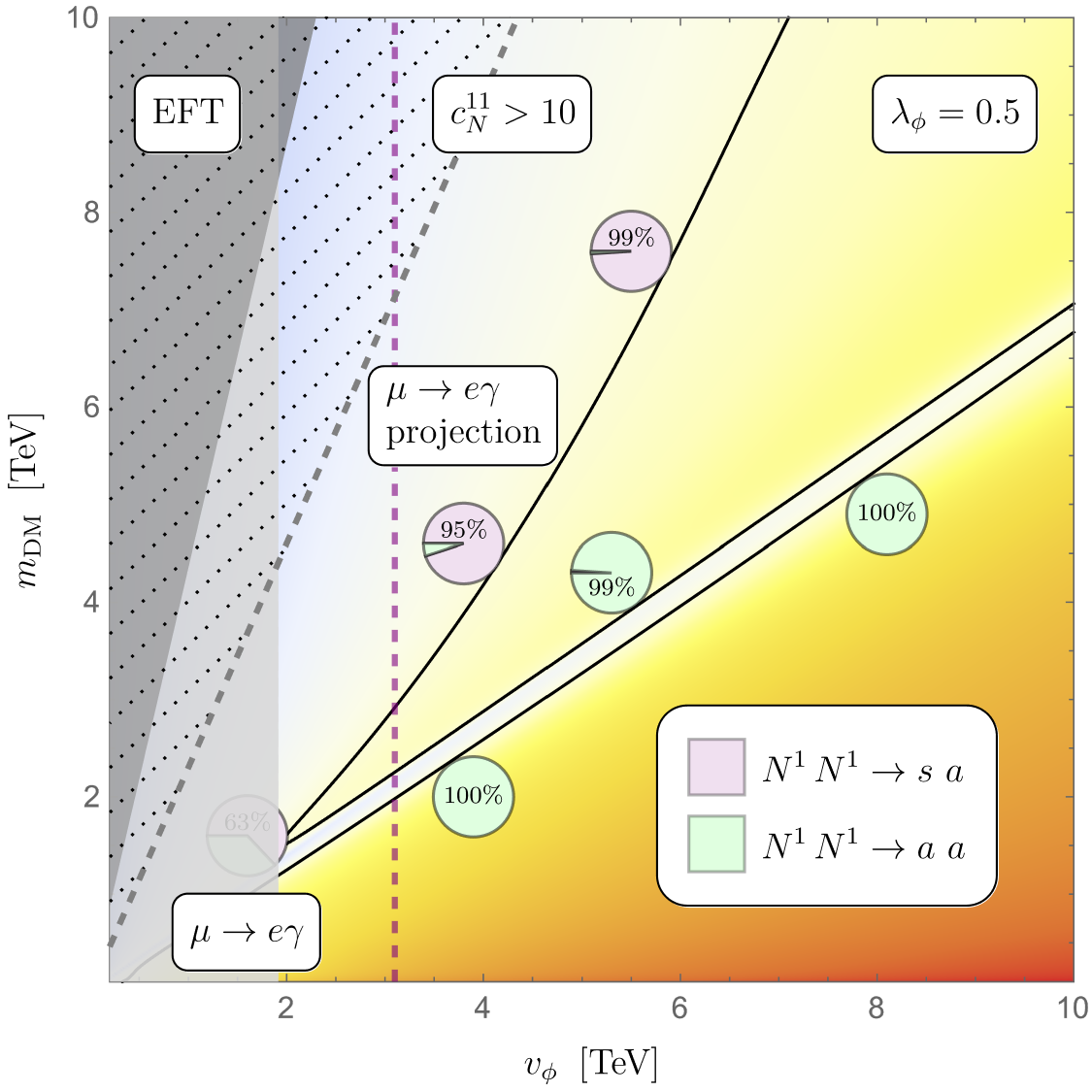}
    \end{subfigure}
    \begin{subfigure}{0.4\textwidth}
        \centering
        \includegraphics[trim={0cm 0cm 0cm 0cm}, clip, width=\linewidth]{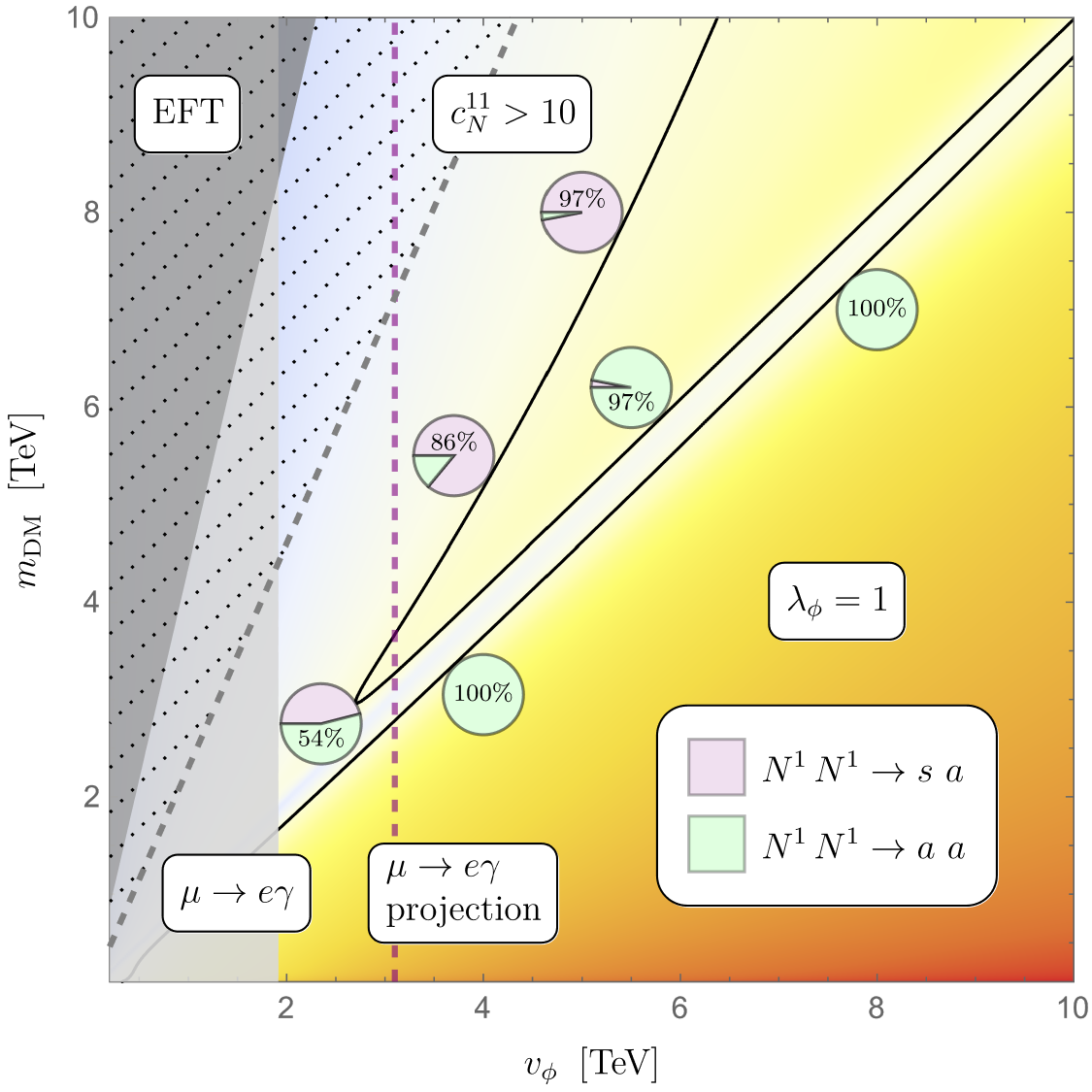}
    \end{subfigure}
    \\
    \begin{subfigure}{0.4\textwidth}
        \centering
        \includegraphics[trim={0cm 0cm 0cm 0cm}, clip, width=\linewidth]{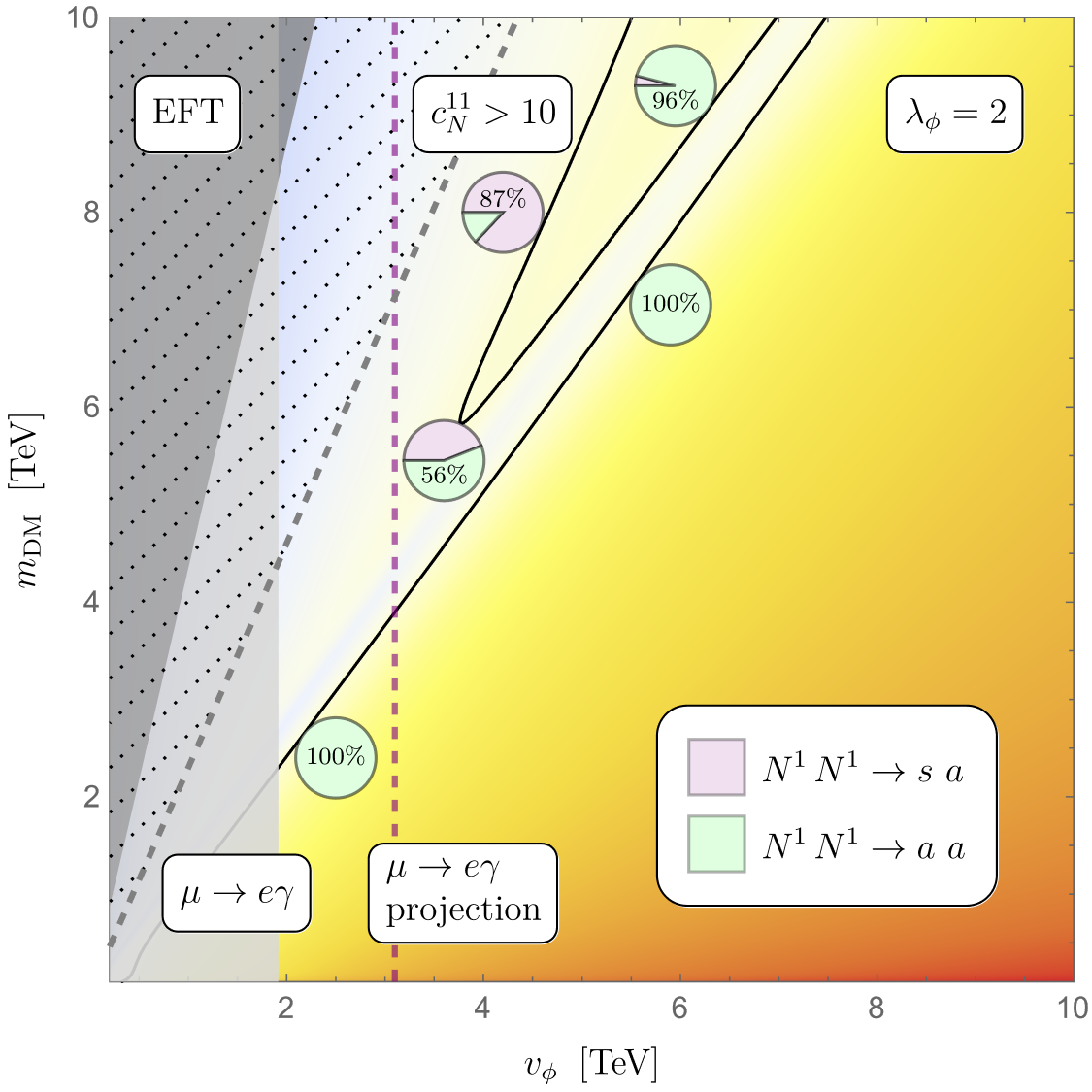}
    \end{subfigure}
    \begin{subfigure}{0.4\textwidth}
        \centering
        \includegraphics[trim={0cm 0cm 0cm 0cm}, clip, width=\linewidth]{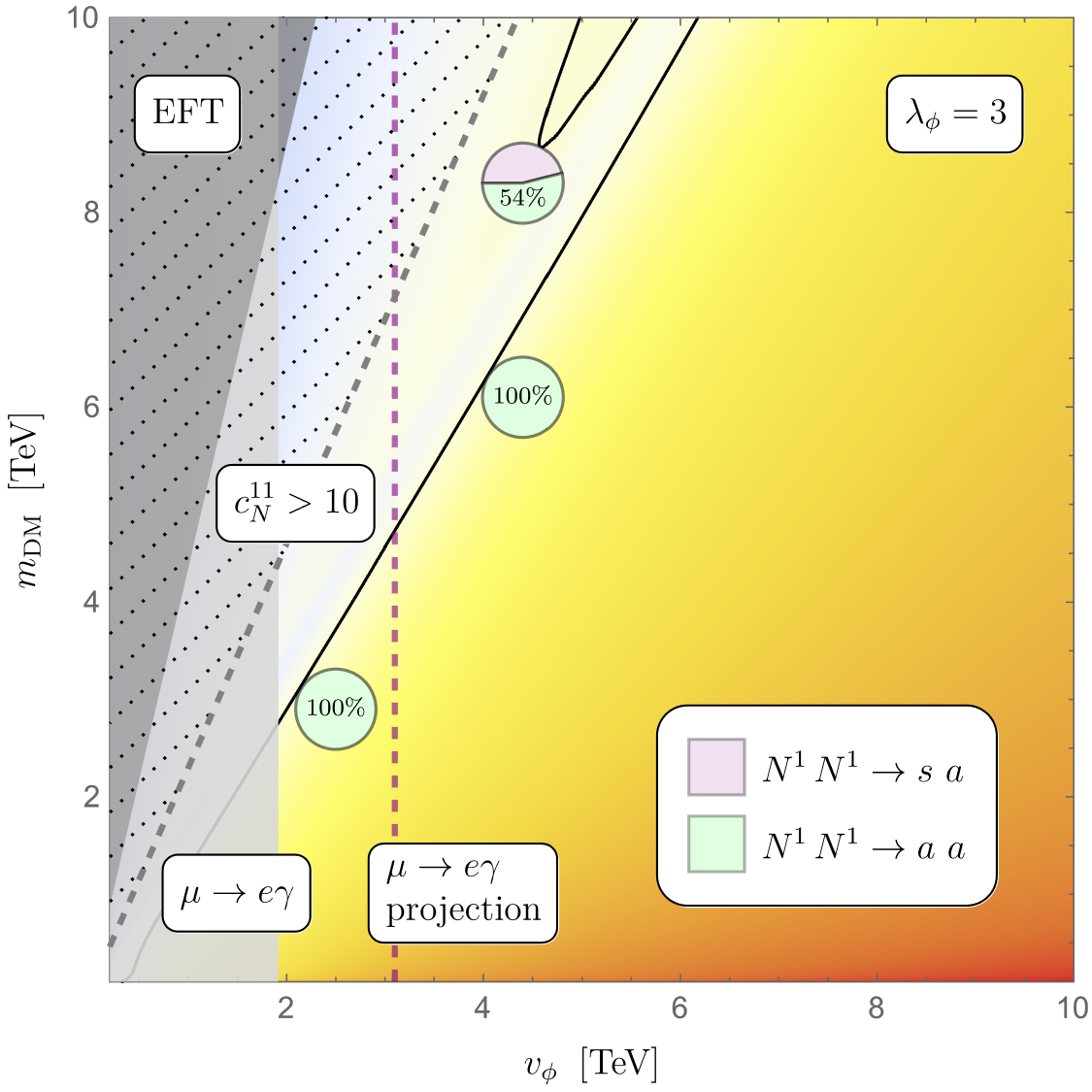}
    \end{subfigure}
    \caption{Similar color code as of Fig.~\ref{fig:freeze-out-1}, with $\lambda_\phi = \{ 0.1,\ 0.3,\ 0.5,\ 1,\ 2,\ 3 \}$, respectively. Black contour lines indicate the observed DM relic abundance $\omg = 0.120 \pm 0.001$~\cite{Planck:2018vyg}. For most of the parameter space in the $v_\phi-\mdm$ plane, a viable freeze-out scenario can be achieved with a suitable value of $\lambda_\phi \in\,[0.1-3]$.}
    \label{fig:diff-lam}
\end{figure}

For example, in the first panel, where $\lambda_\phi = 0.1$, the scalar mass is much lighter, $m_s = 2 \sqrt{\lambda_\phi} \, v_\phi = 0.63 \, v_\phi$. As a result, the \textit{sa-branch} merges with the \textit{aa-branch} at low values of $\mdm$ and $v_\phi$, while $N^1$ annihilating to the SM fermion final states, $N^1 N^1 \to \text{SM}$, contributes substantially to the freeze-out process. However, such low values of $\mdm$ and $v_\phi$ are ruled out by the branching fraction of $\mu \to e \gamma$. Moreover, for $\lambda_\phi = 0.1$ and 0.3, the scalar mass is light enough such that the process $N^1 N^1 \to s \, s$ is no longer kinematically suppressed. Although this contribution is smaller than that of $N^1 N^1 \to s \, a$ but is still visible, as indicated in cyan on the pie chart in \textit{sa-branch}.

The six panels in Fig.~\ref{fig:diff-lam} together demonstrate that the ensemble of solutions of $\mdm$ and $v_\phi$ to achieve the genesis of DM through the freeze-out mechanism covers most of the parameter space we considered. These results support RHN $N^1$ in our model as a viable candidate for DM in the freeze-out scenario.

\subsection{Freeze-in scenario}
\label{sec:freeze-in}

In case the FN symmetry breaking scale $v_\phi$ is sufficiently high,
interactions between the SM and the DM candidate $N_1$ are feeble, and $N_1$ was never in equilibrium with the thermal bath in the early Universe. They can be produced through interactions and decay of flavon particles, 
and the observed relic abundance of DM can be generated through the freeze-in mechanism. In this framework, the particles in the thermal bath annihilate into $N^1$ via the mediator flavon. The process will eventually stop once the thermal bath cooled to a temperature much lower than the mass $N_1$, fixing the comoving number density of $N_1$ particle at constant, namely, `frozen in'. We recall that after spontaneous breaking of the FN symmetry, both the scalar $s$ and the pseudoscalar $a$ components of the flavon field are generated where the pseudoscalar is much lighter than the scalar. Our main focus in this study is to achieve freeze-in via the pseudoscalar portal. Thus we assume that the reheating temperature $\TR$ of the Universe is below the scalar mass $m_s$ but much higher than the pseudoscalar $m_a$ so that the scalar does not bring $N_1$ into equilibrium through its decay. In this way we can isolate the pseudoscalar portal and also draw a consequence in the DM genesis from the reheating of the Universe. An important observation we made in relation to the reheating temperature is the interplay between the contributions of the $2 \to 2$ and $2 \to 3$ processes, which we will discuss later in Sec.~\ref{sec:IR} and~\ref{sec:UV}.

Interestingly, since all the interaction strengths between the flavon and the SM as well as $N^1$ depend on the FN symmetry breaking scale $v_\phi$, we can make an estimate of $v_\phi$ ensuring an out-of-equilibrium nature. Requiring the dominant production rate of all these new physics particles to remain smaller than the expansion rate of the Universe, $\Gamma < H(T) \sim T^2/M_{pl}$ at temperature $T$, we find
\begin{align}
\Gamma_{f\bar{f}^\prime \to a} \sim \frac{1}{8 \pi} \left(\frac{1}{v_\phi}\right)^2 m_a < H(T)\big|_{T=m_a} &\rightarrow v_\phi > 10^8 \, \gev \\
\Gamma_{f\bar{f}^\prime \to N^1 N^1} \sim \frac{1}{4 \pi} \left(\frac{1}{v_\phi}\right)^4 T < H(T)\big|_{T=\mdm} &\rightarrow v_\phi > 10^4 \, \gev \\
\Gamma_{a a \to N^1 N^1} \sim \frac{1}{4 \pi} \left(\frac{1}{v_\phi}\right)^4 T < H(T)\big|_{T=m_a} &\rightarrow v_\phi > 10^4 \, \gev \,,
\end{align}
where we used the reduced Planck mass $M_{pl}= 2.44 \times 10^{18}\, \gev$, $m_a\sim 10\, \gev$, $\mdm \sim\, 100\,\gev$.
Therefore, it is clear from the above estimates that to achieve thermal freeze-in of $N^1$, by producing the pseudoscalar that remains in thermal equilibrium with the thermal bath of SM, requires $10^4 < v_\phi < (10^8 - 10^9)\,$GeV (a similar observation was made in  Fig.~5 in Ref.~\cite{Bharucha:2022lty}).

As mentioned, opposite to the freeze-out scenario, for freeze-in, the SM fermions in the thermal bath annihilate into $N^1$, $\bar{f_i} f_j \to N^1 N^1$ through the flavon portal (see Fig.~\ref{fig:NN-ff} for the Feynmann diagram of the reverse process).
The scalar-mediated channel is suppressed because $s$ is much heavier. 
Both the couplings $\gff$ and $\gdm$ (introduced in Eqs.~\eqref{eq:g+}-\eqref{eq:gDM}) are suppressed by the FN symmetry-breaking scale $v_\phi$ and proportional to the corresponding fermion masses. The cross section of the $\bar{f_i} f_j \to N^1 N^1$ process is proportional to $\gff^2 \times \gdm^2$.
If the pseudoscalar is in thermal equilibrium with the SM, they will dominantly produce $N_1$ through the $t$-channel annihilation $a \, a \to N^1 N^1$ with a larger cross-section,
\begin{align}
\label{eq:aaNN}
\sigma_{aa \to NN}^{t-channel} \ \sim \ \gdm^4 \ > \ \gff^2 \times \gdm^2 \ \sim \ \sigma_{ff \to NN} \,
\end{align}

As a benchmark case that avoids phenomenological constraints, we choose the mass of the pseudoscalar $m_a = 10\,$GeV. Note that the constraints from $\mu \to e \gamma$ and $B$-meson mixing are indeed satisfied for the freeze-in scenario due to the relatively large value of $v_\phi$. The total cross section for the $a \, a \to N^1 N^1$ process in the center-of-mass frame is given in Appendix~\ref{app:Xsec}. The thermally averaged cross section is obtained by Eq.~\eqref{eq:sigmav}, in which $\sigma(E_\text{CM})$ is replaced by Eq.~\eqref{eq:aaNN}. We then calculate the DM yield and the relic density in Sec.~\ref{sec:IR} and~\ref{sec:UV}, with respect to different ranges of reheating temperature, which are usually determined by unknown inflation mechanisms.

\subsubsection{IR freeze-in}
\label{sec:IR}

For sufficiently low $\TR$, the freeze-in is dominated by the $2 \to 2$ process of $a \, a \to N^1 N^1$ generated by renormalizable interactions in Eq.~\eqref{eq:gDM}. Consequently, the DM yield is independent of $\TR$. Such cases are generally known as ``IR freeze-in'' in literature~\cite{Hall:2009bx}, compared to ``UV freeze-in'', which we discuss in Sec.~\ref{sec:UV}.

To obtain the IR freeze-in relic density, we define the dark-matter yield $Y = n_\text{$N^1$} \big/ \se$, where $n_\text{$N^1$}$ is $N_1$ number density and $\se$ is the total entropy density given by
\begin{equation}
\se = \frac{2 \pi^2}{45} \, g_{*s} \, T^3 \,.
\end{equation}
where $g_{*s}$ is the effective degrees of freedom in entropy, for which we use the data in \cite{Husdal:2016haj} for the SM contribution and neglect those of the new physics particles because they are above the GeV range.

The DM $N^1$ yield today $\YIR$ can be calculated by solving the Boltzmann equation. We use the results in \cite{Bharucha:2022lty} for the freeze-in process,
\begin{equation}
\label{eq:Y0IR}
\YIR = - \int_{0}^\infty\frac{\sv \, n_\text{eq}^2}{3H \se^2} \left( \frac{\d \se}{\d T} \right) \d T \,,
\end{equation}
where $\sv$ is the thermally averaged cross section in Eq.~\eqref{eq:sigmav}, $n_\text{eq}$ is the equilibrium number density of $N^1$. Assuming Maxwell-Boltzmann statistics for a given particle of mass $m$ and degrees of freedom $g_\text{dof}$, their equilibrium number density is given by
\begin{align}
n_{\rm{eq}}(T) &= \frac{g_\text{dof}}{(2\pi)^3}\,\int e^{-E/T}\,d^3 p = \frac{g_\text{dof}}{2 \pi ^2}\,m^2\,T K_2\left(m/T\right) \,.
\label{eq:MB}
\end{align}
The $H$ in Eq.~\eqref{eq:Y0IR} is the Hubble rate given by
\begin{align}
H^2 = \frac{8 \pi G}{3} \rho \,,
\end{align}
with the SM energy density given by 
\begin{equation}
\rho = \frac{\pi^2}{30} \, g_{*\rho} \, T^4 \,,
\end{equation}
where $g_{*\rho}$ is the effective degrees of freedom in energy \cite{Husdal:2016haj}. As a candidate for DM, the $N_1$ relic density today can be derived from its yield $\YIR$,
\begin{align}
\omg = \frac{\rho_\text{$N^1$}}{\rho_\text{crit}/h^2} = \frac{\mdm}{\rho_\text{crit}/h^2} \, n_\text{$N^1$} = \frac{\mdm}{\rho_\text{crit}/h^2} \, s_0 \, \YIR \,,
\end{align}
with the critical density $\rho_\text{crit}/h^2=1.053672(24) \times 10^{-5} \ \rm{GeV}\ \rm{cm}^{-3}$, and the entropy density today $s_0=2891.2 \ \text{cm}^{-3}$ \cite{ParticleDataGroup:2022pth}.

\begin{figure}[!ht]
	\begin{center}
		\includegraphics[trim={0cm 0 0 0},clip, width=0.6\textwidth]{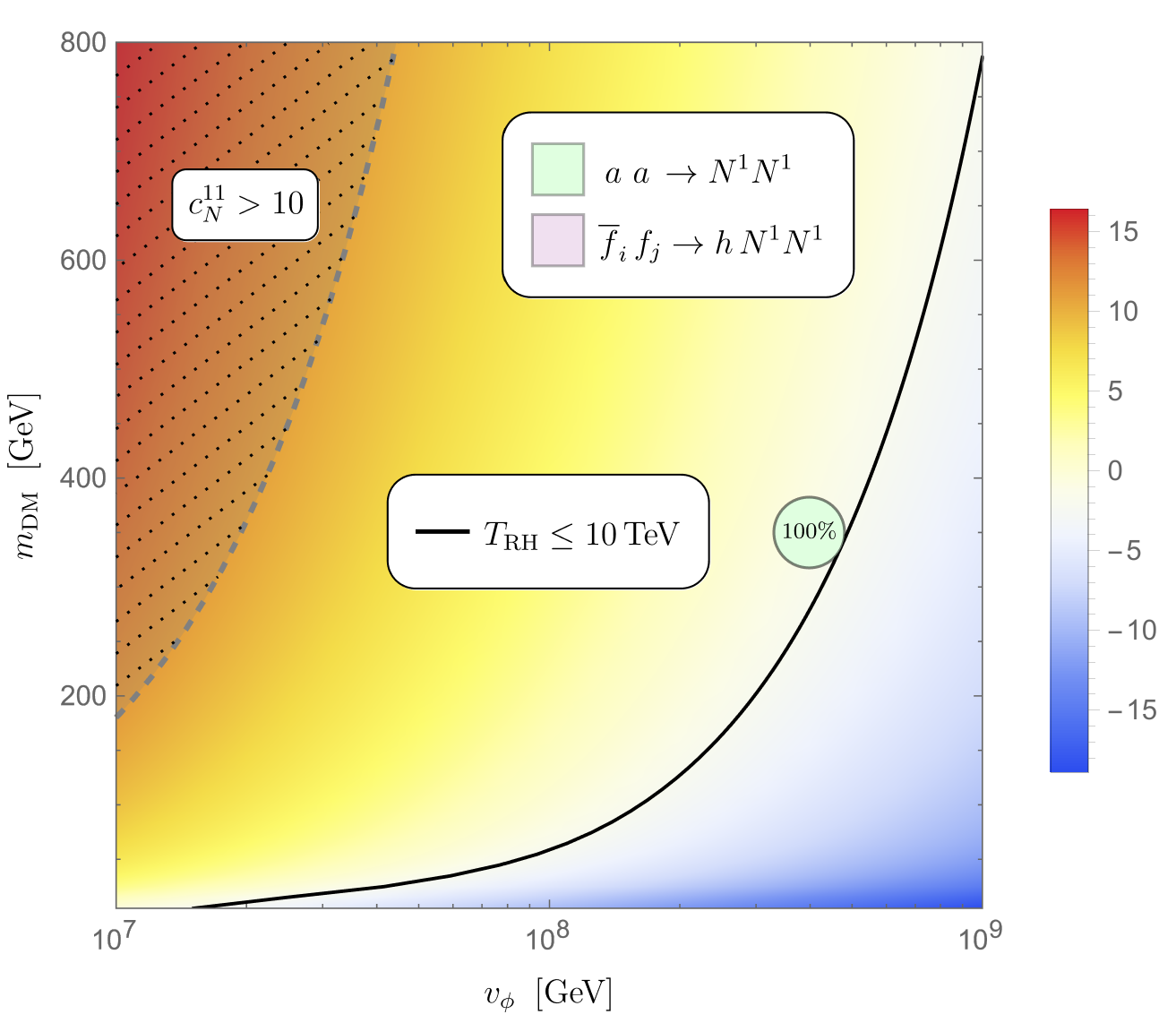}
	\end{center}
	\vspace{-0.5cm}
	\caption{
 Contour plot for the logarithm of relic density, i.e. $\ln \omg$ in the range of $v_\phi \in\,[10^7 - 10^9]\,$GeV and $\mdm \in\,[5-800]\,$GeV for the IR freeze-in scenario with $\TR \leq 10\,$TeV. The black curve indicates the observed DM relic abundance $\omg = 0.120 \pm 0.001$~\cite{Planck:2018vyg}, dominated by the $2 \to 2$ process $a \, a \to N^1 N^1$. The dashed light gray region is excluded because it indicates $c_N^{11} > 10$, which is disfavored by the FN framework.
	}\label{fig:IR-Freeze-in}
\end{figure}

For the observed relic density of the Universe, the result of the parameter scan in the $\mdm - v_\phi$ plane is illustrated in Fig.~\ref{fig:IR-Freeze-in}. Our main focus in this analysis is in the heavy DM mass region, where $2 \to 2$ annihilation processes are the dominant production channels for the DM, rather than from the decay of the pseudoscalar to two DM particles, $a \to N^1 N^1$. We start the scan at $\mdm = m_a / 2 = 5\,$GeV, for which the observed relic abundance is obtained at $v_\phi \sim 10^7\,$GeV. Then we stop the scan at $v_\phi = 10^9\,$GeV to ensure that the pseudoscalar is always in the thermal bath. In this entire range, the observed relic abundance $\omg = 0.12$ can be achieved with $\mdm$ ranging from 5 GeV to 800 GeV, as shown by the black curve in Fig.~\ref{fig:IR-Freeze-in}. The regions on the left and right sides of the black curve represent overabundance and underabundance, respectively. The light gray region to the left of the dashed curve is excluded because it indicates $c_N^{11} > 10$, which is disfavored by the FN framework. Our results show that the IR freeze-in is almost 100\% dominated by the contribution of the $2 \to 2$ process $a \, a \to N^1 N^1$ as long as $\TR \leq 10\,$TeV and high enough to produce DM from the thermal bath.  On the other hand, if $\TR \geq 100\,$TeV, then the UV freeze-in process starts to become important, which we discuss next.

\subsubsection{UV Freeze-in}
\label{sec:UV}

\begin{figure}[H]
\begin{center}
\includegraphics[trim={0cm 0cm 0cm 0cm },clip, width=0.35\textwidth]{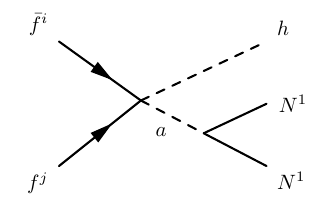}
\end{center}
\vspace{-0.5cm}
\caption{The Feynman diagram for $\bar{f}^i f^j \to h N^1 N^1$ process induced by dimension-five operator.}
\label{fig:ff-hNN}
\end{figure}

We noted in Eq.~\eqref{eq:Lagdim5}, the effective theory expansion in flavon fields induces non-renormalisable interactions through dimension-five operators. They give rise to contact interactions between the SM fermions, the Higgs boson, and the pseudoscalar flavon. These interactions induce $2 \to 3$ processes  $\bar{f}^i f^j \to h N^1 N^1$, see the Feynman diagram in Fig.~\ref{fig:ff-hNN} for details. The $2 \to 3$ processes are UV dominated, which means their contribution to the freeze-in scales as the reheating temperature $\TR$. For sufficiently high $\TR$, freeze-in is dominated by these $2 \to 3$ processes involving non-renormalisable operators. This is usually called the UV freeze-in scenario, which has been studied in the literature~\cite{Elahi:2014fsa, Biswas:2019iqm}.

As mentioned earlier, to focus on the pseudoscalar portal, we assume that $\TR$ is below the scalar mass $m_s$, but much higher than the masses of the rest of the particles. Then the DM yield of the $2 \to 3$ processes $\bar{f}^i f^j \to h N^1 N^1$ can be estimated as~\cite{Elahi:2014fsa}
\begin{align}
    \YUV \approx \frac{135 \, M_{\text{Pl}}}{1.66 \times (2\pi)^9 \, g_{*s} \sqrt{g_{*\rho}}} \left( \frac{ \gff \times \gdm}{v_\text{EW}} \right)^2 \TR \,,
\label{eq:Y0UV}
\end{align}
which is linearly proportional to $\TR$. By comparing this $\YUV$ yield with the $\YIR$, calculated in Eq.~\eqref{eq:Y0IR}, we can estimate the range of $\TR$ where $\YUV$ starts to become important. For example, consider the benchmark point $v_\phi = 5 \times 10^8\,$GeV and $\mdm = 350\,$GeV in Fig.~\ref{fig:IR-Freeze-in}, the observed relic density can be achieved by IR freeze-in. However, we find that if the choice for $\TR$ is such that $\TR > 2.2 \times 10^6\,$GeV, the UV freeze-in yield becomes greater than that of IR, that is, $\YUV > \YIR$ and should not be neglected in the analysis.

\begin{figure}[!h]
	\begin{center}
		\includegraphics[trim={0cm 0 0 0},clip, width=0.5\textwidth]{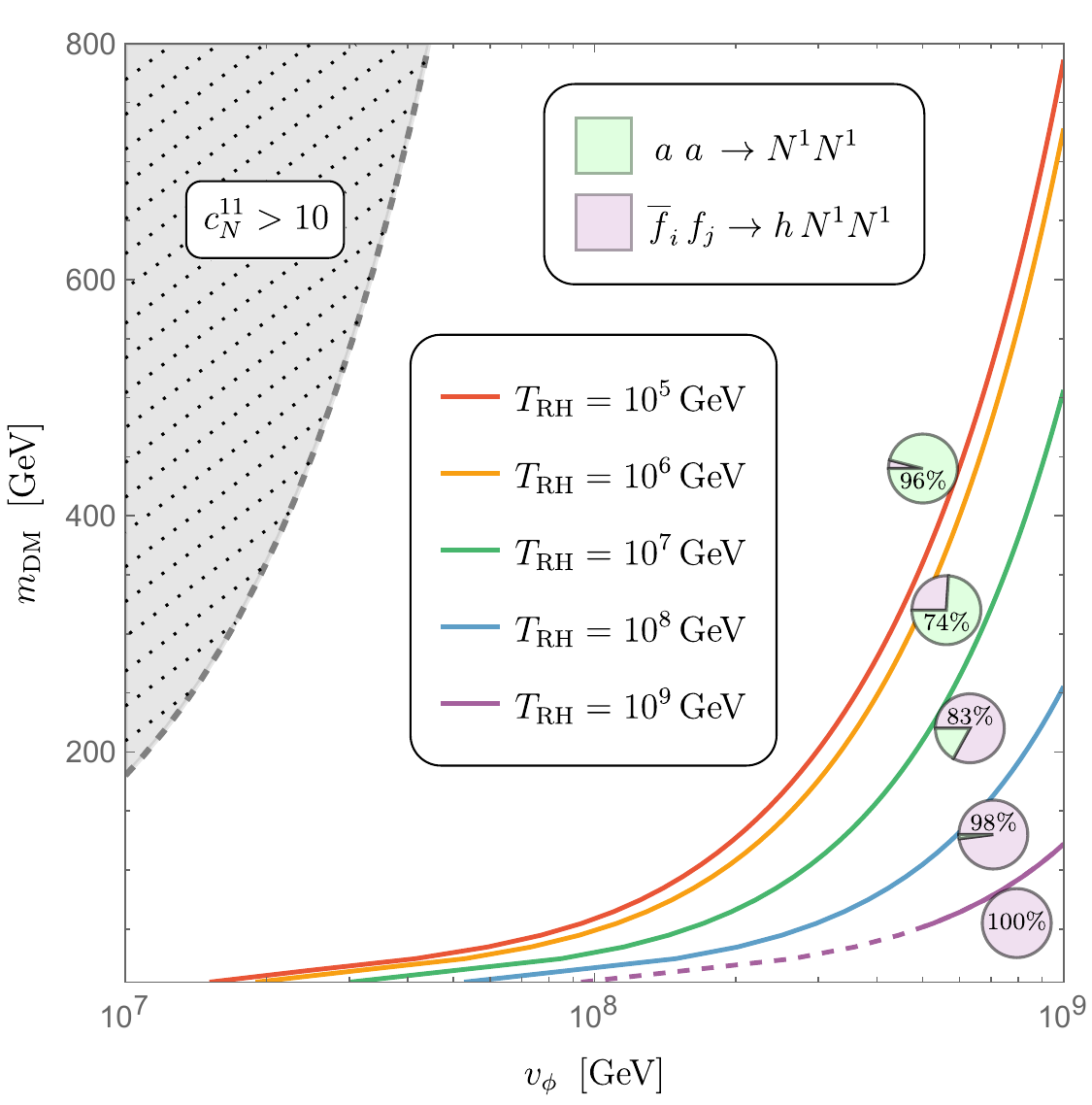}
	\end{center}
	\vspace{-0.5cm}
	\caption{
 Contour plot for the logarithm of relic density, i.e. $\ln \omg$ in the range of $v_\phi \in\,[10^7 - 10^9]\,$GeV and $\mdm \in\,[5-800]\,$GeV for the UV freeze-in scenario with several different choices of $\TR$ from $10^5$ to $10^9\,$GeV. Colored curves represent where $\omg = 0.120 \pm 0.001$~\cite{Planck:2018vyg} is achieved for different $\TR$ values. The pie charts illustrate, in percentage, the contributions from the $2 \to 2$ and $2 \to 3$ processes, specifically at the point where the colored curve intersects the pie chart. The dashed light gray region is excluded because it indicates $c_N^{11} > 10$, which is disfavored by the FN framework. The dashed part of the purple curve is not valid because the scalar mass $m_s = 2 v_\phi$ is supposed to be higher than $\TR$.
}\label{fig:UV-Freeze-in}
\end{figure}

We show the results of UV freeze-in for $\TR$ ranging from $10^5$ to $10^9\,$GeV in Fig.~\ref{fig:UV-Freeze-in}.  Curves in different colors represent the region where the observed relic density is achieved for different values of $\TR$, as mentioned in the legend. For $\TR = 10^9\,$GeV, the dashed part of the purple curve corresponds to $v_\phi < 5 \times 10^8\,$GeV. It is not valid in the particular setup where we have assumed that the scalar mass $m_s = 2 \sqrt{\lambda_\phi} \, v_\phi$ is higher than $\TR$.
The pie charts illustrate in percentage the contributions of the $2 \to 2$ and $2 \to 3$ processes, specifically at the point where the color curve intersects the pie chart. We see that the UV contribution from $\bar{f}^i f^j \to h N^1 N^1$ starts to become substantial for $\TR \sim 10^6\,$GeV, and dominates over the IR freeze-in for $\TR \gtrsim 10^7\,$GeV. Note that the $Y_0^{UV}$ calculated in Eq.~\eqref{eq:Y0UV} is an estimate, and $\mathcal{O}(1)$ corrections to the UV contributions are possible due to uncertainties associated with our model parameters, such as $\gff$, as well as unknown details of the inflation and reheating mechanism, which is beyond the scope of this work.

\section{Direct detection}
\label{sec:detection}

Direct detection is one of the cornerstones of DM searches, especially in probing the weekly interacting massive particle paradigm. While current experiments such as XENONnT~\cite{XENON:2023sxq}, LUX-ZEPLIN (LZ)~\cite{LZ:2022ufs} have already probed the majority of the parameter space of popular beyond standard model theories, the next generation such as DARWIN~\cite{DARWIN:2016hyl}, and future upgrades of these nuclear recoil experiments,  are expected to make substantial progress in sensitivity in the next decade. In this section, we discuss the prospects for the direct detection of our DM candidate in the freeze-out scenario through spin-independent and spin-dependent channels mediated by the scalar and pseudoscalar flavons. The parameter space we obtained for the freeze-in scenario is characterized by quite high $v_\phi$ values, implying tiny couplings to SM fermions and therefore are very challenging to test even in future direct-detection experiments.

\subsection{Spin-independent cross section}
\label{sec:SI}

\begin{figure}[H]
\begin{center}
\includegraphics[trim={0cm 0cm 0cm 0cm },clip, width=0.85\textwidth]{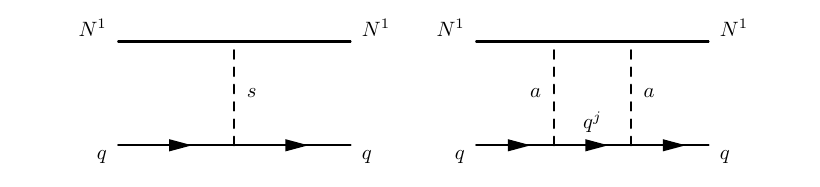}
\end{center}
\vspace{-0.5cm}
\caption{Feynman diagrams for spin-independent scattering of $N^1$ off the SM quarks. (Left): tree-level interaction via a scalar exchange. (Right): box diagram mediated by two pseudoscalars.}
\label{fig:box}
\end{figure}

In this section, we start with direct detection through spin-independent scattering cross-section channels. Such interactions are mediated by a tree-level exchange of the scalar flavon or with two pseudoscalars in the box diagram, as shown in the left and right panels of Fig.~\ref{fig:box}, respectively.

Although the box contribution suffers from an extra loop factor compared to the tree level one, we find the spin-independent amplitudes mediated by the scalar and the box diagram have the same parametric suppression as 
\begin{alignat}{2}
    \text{tree-level:}& &\quad \frac{g_{DM} \times (g_+^f)_{ii}}{m_s^2} \quad &\sim \quad \frac{1}{v_\phi^4} \,, \nn\\[1ex]
    \text{box diagram:}& &\quad g_{DM}^2 \times (g_+^f)_{ij}^2 \quad &\sim \quad \frac{1}{v_\phi^4} \,,
\end{alignat}
where $f=u,\,d$ and $i,\,j = 1,\,2,\,3$, and $m_s = 2 \sqrt{\lambda_\phi} \, v_\phi$. The couplings $g_{DM}$ and $g_+^f$ are suppressed by one power of $v_\phi$. Consequently, both spin-independent amplitudes are suppressed by $v_\phi^4$, compared with $v_\phi^2$ for the spin-dependent amplitudes, which we will see in the next section. Moreover, spin-independent scatterings do not suffer from momentum suppression. In contrast, due to the coherent cross section, an enhancement proportional to the square of the atomic mass $A^2$ further increases the sensitivity of spin-independent direct detection experiments.

In principle, one should combine the tree-level scattering and the box diagram together to calculate the total cross section. However, as we shall see below, they depend on different masses and are subject to different experimental constraints. To better understand these features with the underlying physics, we calculate them separately.
\\
\\
\textbf{Tree-level scalar-mediated scattering:} In the non-relativistic approximation, it can be described by an effective operator as
\begin{align}
\frac{g_{DM} \, g_+^q}{m_s^2} \ \overline{N^1} N^1 \ \overline{q} q\,. 
\end{align}
Here $g_+^q=(g_+^f)_{ii}$, with $f=u,\,d$ and $i=1,\,2,\,3$, is the corresponding flavon coupling to the quark bilinear. For example, if $q\equiv s$, we have $g_+^s=(g_+^d)_{22}$.
The amplitude generated by this operator is proportional to the sum of the nuclear matrix element.
\begin{align}
    \mathcal{M} \sim \sum_{q\,=\,{\rm all\,quarks}} g_+^q \, \VEV{n|\overline{q} q|n} \,,
\end{align}
where $n$ stands for nucleon (proton or neutron). The summation runs through all the SM quarks. Among them, the matrix element of the light quarks (up, down, and strange) can be calculated in chiral perturbation theory~\cite{Cheng:1988cz,Cheng:1988im,Gasser:1990ce} as
\begin{align}
    \VEV{n|\bar{q} q|n} = \frac{m_n}{m_q} f_{Tq}^n \,,
\label{eq:fTq}
\end{align}
where $q = u,\ d,\ s$. The coefficients $f_{Tq}^n$ are taken from~\cite{Ellis:2018dmb,Hoferichter:2017olk} and shown in Tab.~\ref{tab:fT}.

\renewcommand{\arraystretch}{1.5}
\begin{table}[ht]
  \centering
  \begin{tabular}{|c|c|c|c|c|}
    \hline
     & $f_{Tu}^n$ & $f_{Td}^n$ & $f_{Ts}^n$ & $f_{TG}^n$ \\ \hline
    Proton & 0.018(5) & 0.027(7) & 0.037(17) & 0.917(19) \\
    Neutron & 0.013(3) & 0.040(10) & 0.037(17) & 0.910(20) \\ \hline
  \end{tabular}
  \caption{Coefficients of matrix elements for quark operators $\VEV{n|\bar{q} q|n}$.}
  \label{tab:fT}
\end{table}

\begin{figure}[ht!]
	\begin{center}
		\includegraphics[trim={0cm 0cm 0cm 0cm },clip, width=0.8\textwidth]{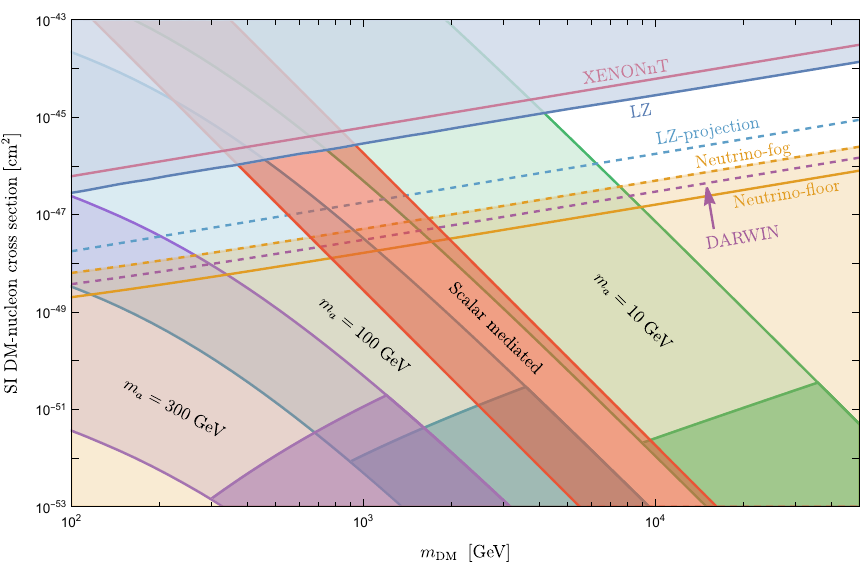}
	\end{center}
	\vspace{-0.5cm}
	\caption{
		The spin-independent DM-Nucleon cross section with the variation of DM mass for both the tree-level scalar-mediated process in the red band and the box-diagram contributions separated for three benchmark values of $m_a=10\,\gev,~100\,\gev$, and 300\,$\gev$ in the green, cyan, and purple bands, respectively. For all bands, the coupling $\lambda_\phi$ is varied between 0.3 to 3. Here $v_\phi$ is estimated to obtain the observed relic density. The darker parts at the bottom of the bands are allowed by the $\mu \to e \gamma $ and $B$-meson mixing data. The blue-shaded region is excluded by LZ~\cite{LZ:2022ufs} and the yellow-shaded area falls within the neutrino-fog~\cite{OHare:2021utq}. Future projections of LZ~\cite{LZ:2018qzl} and DARWIN~\cite{DARWIN:2016hyl} are also shown in dashed blue and dashed purple lines, respectively.
	}\label{fig:DD}
\end{figure}

Note that, as the interaction strengths of flavon particles with SM quarks are proportional to the mass of the quarks, the effect of heavy quarks (top, bottom, and charm) can be substantial and hence should not be neglected. Now, as these heavy quarks are heavier than the nucleons, they should be integrated out, replaced by an effective theory describing nuclear physics. To this end, we relate the heavy-quark content of the nucleus to the gluon condensate~\cite{Shifman:1978zn}
\begin{align}
    m_q \bar{q} q = -\frac{\alpha_s}{12 \pi} G_{\mu\nu} G^{\mu\nu} \,,
\label{eq:Shifman}
\end{align}
where $\alpha_s$ is the strong coupling constant and $G_{\mu\nu}$ is the gluon field strength. Then the matrix elements of the heavy quarks can be approximated through the gluon condensate as
\begin{align}
    \VEV{n|\bar{q} q|n} = -\frac{\alpha_s}{12 \pi \, m_q} \VEV{n|G_{\mu\nu} G^{\mu\nu}|n} = \frac{2}{27} \left( \frac{m_n}{m_q} \right) f_{TG}^n \,,
\label{eq:fTG}
\end{align}
where $q = c,\ b,\ t$, and the coefficient $f_{TG}^n$ is also given in Tab.~\ref{tab:fT}. Combining the contributions from light quarks in Eq.~\eqref{eq:fTq} and those from the heavy quarks in Eq.~\eqref{eq:fTG}, the cross section is given by
\begin{align}
    \sigma_{\text{scalar}}^{\text{SI}} = \frac{4 \, \mu_n^2 \, m_n^2 \, g_{DM}^2}{\pi \, m_s^4} \left| \sum_{q=u,\,d,\,s} \frac{g_+^q}{m_{q}} f_{Tq}^n + \frac{2}{27} \, f_{TG}^n \sum_{q=c,\,b,\,t} \frac{g_+^q}{m_{q}} \right|^2 \,,
\label{eq:sigma_scalar}
\end{align}
where $\mu_n$ is the reduced mass of $N^1$ and the nucleon, and the prefactor 4 is due to $N^1$ being Majorana fermion.  As a result, we compute the spin-independent cross section of the tree-level scalar exchange, and present our results \footnote{Our results agree with those of \texttt{MicrOMEGAs}} as the red band in Fig.~\ref{fig:DD} which will be discussed in detail.
\\
\\
\textbf{Box-diagram pseudoscalar-mediated scattering:} We adapt a formalism similar to that discussed in Refs.~\cite{Arcadi:2017wqi,Freytsis:2010ne}. We take the non-relativistic approach and start with an effective Lagrangian given by \footnote{Another detailed computation can be found in~\cite{Hisano:2010ct}, where the top-mass dependence of the DM-gluon effective interaction is considered.}
\begin{align}
    \mathcal{L} &= g_{DM}^2 \sum_{q} B_{q} \, \overline{N^1} N^1 \ \overline{q} q\,, \\[1ex]
  {\rm where~}  B_{q} &= \sum_{j=1,2,3} \left| (g_+^f)_{ij} \right|^2 C_{S,q^j}(\mdm, \, m_a, \, m_{q^j}) \,.
\label{eq:Lag_SI}
\end{align}
Here we introduced $B_q$ as a short-hand notation.
For example, if $q\equiv s$, we have $(g_+^d)_{2j}$ as the coupling strengths entering $B_q$. The function $C_{S,q^j}$ is derived from the Passarino-Veltman integrals~\cite{Passarino:1978jh}. It has the dimension of inverse mass squared and depends on the mass of the DM, the pseudoscalar, and the quark $q^j$ in the loop. An analytical form of $C_{S,q^j}$ can be found in Appendix A of~\cite{Arcadi:2017wqi}. It is worth mentioning that we neglect vector current operators of the form $\overline{N^1} \gamma_\mu N^1 \ \overline{q} \gamma^\mu q \,$ in the effective Lagrangian because only valence quarks contribute to the nuclear expectation value of the vector current. Nevertheless, $g_+^f$ are generally proportional to the quark masses, which are negligible in the case of valence quarks.

Using Eqs.~\eqref{eq:fTq} to~\eqref{eq:sigma_scalar} for the nuclear matrix element $\VEV{n|\overline{q} q|n}$, we obtain the spin-independent cross section generated by the box diagram,
\begin{align}
     \sigma_{\text{box}}^{\text{SI}} = \frac{4}{\pi} \, \mu_n^2 \, m_n^2 \, g_{DM}^4 \left| \sum_{q=u,d,s} \frac{B_{q}}{m_{q}} f_{Tq}^n + \frac{2}{27} \, f_{TG}^n \sum_{q=c,b,t} \frac{B_{q}}{m_{q}} \right|^2 \,.
\label{eq:sigma_box}
\end{align}

In Fig.~\ref{fig:DD}, we estimate the spin-independent DM-nucleon cross sections over the DM mass range for the freeze-out scenario and compare them with experimental constraints. The red band corresponds to the cross sections generated by the tree-level scalar exchange (Fig.~\ref{fig:box} left panel). For a given DM mass, the values of $v_\phi$ that achieve freeze-out depend on $\lambda_\phi$, while the cross section of the scalar exchange $\sigma_{\text{scalar}}^{\text{SI}}$ depends on both $\lambda_\phi$ and $v_\phi$, as can be seen in Eq.~\eqref{eq:sigma_scalar}. To compute $\sigma_{\text{scalar}}^{\text{SI}}$ for the freeze-out scenario, we scan the parameter space of the DM mass from 100 GeV to 50 TeV and vary $\lambda_\phi$ from 0.3 to 3, resulting in the red band that indicates where the observed relic is achieved.

Regarding the box-diagram contribution (Fig.~\ref{fig:box} right panel), we implement the same scan strategy, varying $\lambda_\phi$ from 0.3 to 3. However, $\sigma_{\text{box}}^{\text{SI}}$ also depends on the mass of the pseudoscalar, $m_a$, for which we choose three benchmark values, 10 GeV, 100 GeV, and 300 GeV, presented in green, cyan, and purple bands, respectively. In all three bands, darker regions at the bottom are \textit{allowed}, while the regions in light shaded color are \textit{excluded} by constraints from low-energy data arising from $\mu \to e \gamma$ and $B$-meson mixing.

We illustrate the current most stringent bounds from direct detection experiments such as XENONnT~\cite{XENON:2023sxq} (in magenta) and LZ~\cite{LZ:2022ufs} (in blue), while the future projections of LZ~\cite{LZ:2018qzl} and DARWIN~\cite{DARWIN:2016hyl} are also shown in dashed blue and dashed purple. The light blue region at the top is excluded by LZ. The light-yellow region covering the lower half of the plot corresponds to the neutrino fog. The solid yellow line represents the traditional neutrino floor taken from the APPEC report~\cite{Billard:2021uyg}. The dashed yellow line draws the boundary of the neutrino fog, which has been discussed in detail in~\cite{OHare:2021utq}.
Compared to the traditional neutrino floor, this definition of neutrino fog does not rely on arbitrary choices of experimental exposure and
energy threshold and, thus, is conservative.

Both the tree-level scalar exchange and the box diagram contribute to the spin-independent scattering. In principle, we should combine their amplitudes and compute the total spin-independent cross section. However, the box diagram depends on the pseudoscalar mass $m_a$ and is constrained by the low energy data, whereas the scalar exchange does not. Therefore, we show the two corresponding cross sections separately to better understand the parametric dependence. In Fig.~\ref{fig:DD}, the scalar exchange (red band) is in between the two box diagram contributions for $m_a = 10\,$GeV (green band) and $m_a = 100\,$GeV (cyan band). In other words, for a 10 GeV pseudosalar, the box diagram overwhelms the scalar exchange and dominates the cross section. On the contrary, for a pseudoscalar at 100 GeV or higher, the scalar exchange supersedes the box-diagram contribution and dominates. Interestingly, for pseudoscalar mass around 30 to 40 GeV, a cancellation between the scalar exchange and the box diagram can happen for DM masses in the range of $\mathcal{O}(100\,\text{GeV})$ -- $ \mathcal{O}(\text{TeV})$, due to the negative sign arising in the loop integral $C_{S,q^j}$.

In summary, the direct detection constraints on the parameter space of our model for $m_a \geq \mathcal{O}(100)\,$GeV come mainly from the scalar exchange, where the current observed limits disfavor DM masses from a few hundred GeV up to 1\,TeV. The actual value of the exclusion depends on $\lambda_\phi$. The next-generation direct detection experiments may further probe the TeV mass range for our DM candidate. On the other hand, if the mass of the pseudoscalar is $\mathcal{O}(10)\,$GeV, then the low-energy data provide the most stringent constraints, excluding DM masses below a few TeV. Moreover, as illustrated by the darker green region, the parameter space allowed for $m_a \sim \mathcal{O}(10)\,$GeV is entirely inside the neutrino fog. Future DM detection experiments with directional sensitivity~\cite{Vahsen:2021gnb,OHare:2020lva} may be able to probe this scenario.

\subsection{Spin-dependent cross section}
\label{sec:SD}
\begin{figure}[H]
\begin{center}
\includegraphics[trim={0cm 0cm 0cm 0cm },clip, width=0.5\textwidth]{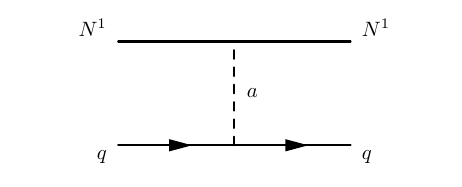}
\end{center}
\vspace{-0.5cm}
\caption{Pseudoscalar flavon mediated tree-level scattering of $N^1$ off SM quarks.}
\label{fig:NN-a-qq}
\end{figure}

As a mediator, the pseudoscalar flavon is much lighter than the scalar. The right-handed neutrino $N^1$ can scatter off nucleons by the exchange of a pseudoscalar flavon, as shown in Fig.~\ref{fig:NN-a-qq}.

In the non-relativistic approximation, this scattering process can be described by an effective operator as
\begin{align}
    \frac{g_{DM} \, g_+^q}{m_a^2} \ \overline{N^1} \gamma_5 N^1 \ \overline{q} \gamma_5 q \ ,
\end{align}
Previous studies have pointed out that the cross section generated by this type of operator is not only spin dependent, but also momentum suppressed, that is, $\sigma \sim k^4$, where $k \sim 100\,$MeV is the typical momentum for the interaction~\cite{Freytsis:2010ne,Fan:2010gt,Agrawal:2010fh}.

To calculate the amplitude, we need to invoke the nuclear matrix elements induced by the quark-level pseudoscalar couplings,
\begin{align}
    g_+^q \ \overline{q} \gamma_5 q \quad \equiv \quad g_n \, m_n \, \overline{n} \gamma_5 n \,,
\end{align}
where $n$ stands for nucleon (proton or neutron). The coupling $g_n$ can be calculated by relating the pseudoscalar interaction to the axial current through PCAC \footnote{PCAC stands for Partial Conservation of Axial Current.} or the generalized Goldberger-Treiman relations~\cite{Fan:2010gt}. Taking the divergence of the axial current and implementing the equation of motion~\cite{Cheng:1988im,Cheng:2012qr}, we obtain
\begin{align}
    g_n &= \sum_{q=u,d,s} \Big( g_+^q - \bar{m} \, \xi \Big) \, \frac{\Delta_q^n}{m_q} \,, \\[1ex]
    {\rm where} \quad \bar{m} &= \Big( m_u^{-1} + m_d^{-1} + m_s^{-1} \Big)^{-1} \quad {\rm and} \quad \xi = \sum_{q\,=\,\text{all\,quarks}} \frac{g_+^q}{m_q} \,.
\end{align}
Here $\Delta_q^n$ is the fraction of the spin of the nucleon carried by the light quarks, and the values are quoted in Table~\ref{tab:Delta}.
%
\renewcommand{\arraystretch}{1.5}
\begin{table}[ht]
  \centering
  \begin{tabular}{|c|c|c|c|}
    \hline
     & $\Delta_u^n$ & $\Delta_d^n$ & $\Delta_s^n$ \\ \hline
    Proton & 0.80(3) & -0.46(4) & -0.12(8) \\
    Neutron & -0.46(4) & 0.80(3) & -0.12(8) \\ \hline
  \end{tabular}
  \caption{Fraction of the spin of the nucleon carried by the light quarks. Values are taken from Ref.~\cite{Hill:2014yxa}.}
  \label{tab:Delta}
\end{table}

As mentioned above, the scattering process mediated by the pseudoscalar is momentum suppressed. To calculate the cross section, we take the non-relativistic approximation~\cite{Freytsis:2010ne}
\begin{align}
    \overline{N^1} \gamma_5 N^1 \quad &\sim \quad \frac{|\Vec{k}|}{2\mdm} \, \overline{N^1} \gamma^\mu \gamma_5 N^1 \nn\\[1ex]
    \overline{n} \gamma_5 n \quad &\sim \quad \frac{|\Vec{k}|}{2m_n} \, \overline{n} \gamma^\mu \gamma_5 n \,,
\end{align}
where we choose the momentum $|\Vec{k}| \sim 100\,$MeV, as a typical value for most spin-dependent detectors. Finally, the spin-dependent cross section is given by
\begin{align}
    \sigma^{\text{SD}} = \frac{3 \, \mu_n^2 \, g_n^2 \, k^4}{2\pi \, v_\phi^2 \, m_a^4} \,.
\end{align}

For our parameter space of DM freeze-out, we estimate the spin-dependent cross section of $N^1$ scattering off a nucleon to be
\begin{align}
    \mqty[ \sigma_p^{\text{SD}} \\[1ex] \sigma_n^{\text{SD}} ] \approx \mqty[ 10^{-46} \\[1ex] 10^{-47}] \, \text{cm}^2 \times \left( \frac{100 \, \text{GeV}}{\mdm} \right)^2 \left( \frac{100 \, \text{GeV}}{m_a} \right)^4 \,,
\label{eq:SD}
\end{align}
where $p$ and $n$ denote proton and neutron, respectively.

The most stringent direct detection constraints to date on the spin-dependent WIMP-nucleon cross section come from experiments such as LZ~\cite{LZ:2022ufs}, XENONnT~\cite{XENON:2023sxq}, and PICO-60~\cite{PICO:2019vsc}. In the DM mass range for our freeze-out scenario, the spin-dependent constraints on the DM-proton interaction are around $10^{-41} \, \text{cm}^2$, while $10^{-42} \, \text{cm}^2$ for the DM-neutron interaction. These bounds are at least five orders of magnitude above the cross section we obtain for our model in Eq.~\eqref{eq:SD}. Therefore, we conclude that current bounds from spin-dependent DM direct detection experiments do not exert constraints on the parameter space of our model. However, improvements in the sensitivity in future upgrades may probe the region.

\section{Light neutrino mass generation}
\label{sec:neutrino}

In this section, we discuss the framework to generate the tiny masses of the light neutrinos via the Type-I seesaw mechanism.  In our model, we have two right-handed neutrinos $N^2$ and $N^3$ that interact with left-handed neutrinos through the SM Higgs doublet, while $N^1$ serves as the DM candidate, as shown in Eq.~\eqref{eq:Lag}. We see in the following that since the DM mass depends on the FN charge of $N^1$, to generate the desired mass spectrum along with the light neutrino masses and mixing patterns, suitable choices of FN charges for right-handed neutrinos are necessary. After introducing the Type-I seesaw mechanism, we discuss the two scenarios, namely DM freeze-out and freeze-in, in the context of FN charge assignments for the lepton sector. \\

Including the $U(1)_\text{FN}$ charges of the right-handed neutrinos $q_{N_{1,2,3}}$, the leptonic part of Eq.~\eqref{eq:FNsol} can be written as
\begin{align}
\left(\begin{array}{ccc}
q_{L_{1}} & q_{L_{2}} & q_{L_{3}} \\
q_{N_{1}} & q_{N_{2}} & q_{N_{3}} \\
q_{e} & q_{\mu} & q_{\tau}
\end{array}\right) \,.
\end{align}
From Eq.~\eqref{eq:Lag}, the Dirac and Majorana mass matrices of neutrinos are given by
\begin{align}
m_{\nu D}^{ik} \ &= \ v_\text{EW} \, c_\nu^{ik} \, \epsilon^{(q_{L_i} - \, q_{N_k})} \,,\nn\\[1ex]
m_{\nu M}^{\alpha \beta} \ &= \ M \, c_N^{\alpha \beta} \, \epsilon^{-(q_{N_\alpha} + \, q_{N_\beta})} \,,
\end{align}
where $m_{\nu D}^{ik}$ is a $3 \times 2$ matrix and $m_{\nu M}^{\alpha \beta}$ is a $2 \times 2$ matrix, with $i = 1, 2, 3$ and $k, \alpha, \beta = 2, 3$.

After integrating out the heavy right-handed neutrinos, the light neutrino mass matrix is obtained through the seesaw mechanism,
\begin{align}
m_{\nu}^{ij} = \left(m_{\nu D}\right) \cdot \left(m_{\nu M}\right)^{-1} \cdot \left(m_{\nu D}\right)^T \sim \frac{v_\text{EW}^2}{M} \, \epsilon^{(q_{L_i} + \, q_{L_j})} \,,
\end{align}
where $i, j = 1, 2, 3$. Note that $m_{\nu}^{ij}$ does not depend on the FN charges of the right-handed neutrinos $q_{N_{1,2,3}}$, and it contains a zero eigenvalue because the Majorana mass matrix $m_{\nu M}$ is rank 2, which means that the lightest active neutrino is massless. Rotating to the physical mass basis, $m_{\nu}^{ij}$ is diagonalized as
\begin{align}
m_\nu \sim \frac{v_\text{EW}^2}{M} \begin{pmatrix}
    0 & & \\
      & \epsilon^{2 q_{L_2}} & \\
      & & \epsilon^{2 q_{L_3}}
\end{pmatrix} \,,
\end{align}
and the PMNS matrix is given by
\begin{align}
	U_\text{PMNS} \sim \begin{pmatrix}
		1 &\epsilon^{q_{L_1}-\,q_{L_2}} & \epsilon^{q_{L_1}-\,q_{L_3}} \\
		\epsilon^{q_{L_1}-\,q_{L_2}} & 1 & \epsilon^{q_{L_2}-\,q_{L_3}} \\
		\epsilon^{q_{L_1}-\,q_{L_3}} & \epsilon^{q_{L_2}-\,q_{L_3}} & 1 
	\end{pmatrix} \,.
 \label{PMNS}
\end{align}
Therefore, substantial mixing between the second and third generation of neutrinos can be achieved by setting $q_{L_2}=q_{L_3}$, and smaller mixing between the first and third generation can be achieved by $q_{L_1}=q_{L_3} + 1$.

\subsection{Freeze-out case}
In the freeze-out scenario discussed in Sec.~\ref{sec:freeze-out}, $v_\phi$ is of $\mathcal{O}(1 - 10)\,$TeV. By denoting $m_\nu^{tot}$ as the total mass of three active neutrinos, we find
\begin{align}
\label{eq:nu_mass}
m_\nu^{tot} \ \sim \ \frac{v_\text{EW}^2}{M} \, \epsilon^{2 q_{L_3}} \ = \ \frac{v_\text{EW}^2}{v_\phi} \, \epsilon^{(2q_{L_3}+\,1)} \,.
\end{align}
In view of the cosmological and oscillation experiment bound on $m_\nu^{tot} \lesssim 0.1\,$eV, and with $v_\text{EW} = 174\,$GeV, $\epsilon = 0.23$, and choosing $v_\phi = 5\,$TeV as a reference case for freeze-out, we obtain $q_{L_3} = 8$, indicating $\{ q_{L_1}, \ q_{L_2}, \ q_{L_3} \} = \{ 9, \ 8, \ 8 \}$ a viable solution for FN charges of the lepton doublet that can generate the observed neutrino mass and mixing textures. \\

The mass of DM depends on the FN charge of $N^1$, as
\begin{align}
\mdm \ = \ M \, c_N^{11} \, \epsilon^{n_N^{11}} \ = \ \frac{v_\phi}{\epsilon} \, c_N^{11} \, \epsilon^{-2 q_{N_1}} \,.
\end{align}
In Fig.~\ref{fig:freeze-out-1}, the observed relic density of DM is realized near the resonance of the $s$-channel where $\mdm \sim v_\phi$, implying that
\begin{align}
\label{eq:mDM_qN1}
\frac{\mdm}{v_\phi} \ = \ c_N^{11} \, \epsilon^{-(2 q_{N_1}+1)} \ \sim \ \mathcal{O}(1) \,.
\end{align}
According to the FN framework, $c_N^{11} \sim \mathcal{O}(1)$ would require $q_{N_1} = 0 \ \text{or}~-1$ because $q_{N_i} \leq 0$ for the three generations of RHN (see Eq.~\eqref{eq:n-to-q} for details). Since $N^1$ is the lightest RHN, we expect that its mass will be smaller than $N^2$ and $N^3$, which implies $q_{N_1} < q_{N_2}$ and $q_{N_3}$. Therefore, a viable option is $\{ q_{N_1}, \ q_{N_2}, \ q_{N_3} \} = \{ -1, \ 0, \ 0 \}$. As a result, the FN charge assignment 
\begin{align}
\left(\begin{array}{ccc}
q_{L_{1}} & q_{L_{2}} & q_{L_{3}} \\
q_{N_{1}} & q_{N_{2}} & q_{N_{3}} \\
q_{e} & q_{\mu} & q_{\tau}
\end{array}\right) = 
\left(\begin{array}{ccc}
9 & 8 & 8 \\
-1 & 0 & 0 \\
0 & 3 & 5
\end{array}\right),
\end{align}
will reproduce both lepton Yukawa structures given in Eq.~\eqref{eq:nFN} and the desired phenomenology discussed in the freeze-out scenario considered in this work.

\subsection{Freeze-in case}
In the freeze-in scenario sketched in Sec.~\ref{sec:freeze-in}, the generation of light neutrino masses follows in a similar way as of Eq. \eqref{eq:nu_mass}, although in this case $v_\phi$ is much higher, around $\mathcal{O}(10^8)\,$GeV. Hence, here we obtain $q_{L_3} = 5$, indicating that $\{ q_{L_1}, \ q_{L_2}, \ q_{L_3} \} = \{ 6, \ 5, \ 5 \}$ can be a viable solution for the observed neutrino mass and the mixing pattern.

In order to keep the DM mass in the $\mathcal{O}(100)\,$GeV range, as demonstrated in Fig.~\ref{fig:IR-Freeze-in} for the IR freeze-in, we find $q_{N_1} = -5$ as a solution to Eq.~\eqref{eq:mDM_qN1}. Therefore, the FN charge assignment 
\begin{align}
\left(\begin{array}{ccc}
q_{L_{1}} & q_{L_{2}} & q_{L_{3}} \\
q_{N_{1}} & q_{N_{2}} & q_{N_{3}} \\
q_{e} & q_{\mu} & q_{\tau}
\end{array}\right) = 
\left(\begin{array}{ccc}
6 & 5 & 5 \\
-5 & 0 & 0 \\
-3 & 0 & 2
\end{array}\right),
\end{align}
will reproduce both the lepton Yukawa structures (shown in Eq.~\eqref{eq:nFN}) and the phenomenology required for the IR freeze-in mechanism of DM genesis.

Regarding UV freeze-in with a much higher reheating temperature, for example, the purple curve in Fig.~\ref{fig:UV-Freeze-in} which corresponds to $\TR = 10^9\,$GeV, the DM mass required to achieve freeze-in is much smaller. In this case, setting $q_{N_1} = -6$ will provide the extra suppression on $\mdm$, which is crucial for a viable solution within the FN framework.

\section{Summary and discussion}
\label{sec:summary}

We investigate the extension of the Standard Model via a complex scalar field, known as a flavon, responsible for the spontaneous breaking of a global $U(1)$ symmetry where all the Standard Model fermions are charged under it. The mass and mixing pattern of quarks and leptons are generated in orders of the vacuum expectation value of the flavon field $v_\phi$ divided by the cutoff scale of the theory through the Froggatt-Nielsen mechanism. We included three right-handed neutrinos in the model where the lightest of them can serve as a dark matter candidate and the other two generate masses for the Standard Model neutrinos through the Type-I seesaw mechanism. 
In this minimal setup, the interaction between the dark and the Standard Model sectors is mediated by both the scalar and pseudoscalar components of the flavon. The natural choice with $\mathcal{O}(1)$ coupling makes the scalar significantly heavier than the pseudoscalar, which is a pseudo-goldstone boson of spontaneous breaking of the $U(1)$ symmetry, and thus we focus more on the pseudoscalar portal in this analysis.

The predictive nature of such theories with dynamical generation of the Yukawa structure of the Standard Model largely fixes the interaction strength between the flavon and fermion fields in terms of only one parameter $v_\phi$. First, listing the possible dominant constraints on $v_\phi$ arising mainly from the meson mixing, radiative lepton decay and direct search results at the LHC,  we explore the thermal history of the dark matter candidate both via the freeze-out and freeze-in mechanisms. It is expected that as the standard freeze-out mechanism assumes that dark matter and mediators are in thermal equilibrium, the scale of $v_\phi$ remains low, $\mathcal{O}(\tev)$, in order to obtain the observed relic density of the Universe. We find the dominant channels are the pair annihilation of dark matter to (pseudo)scalar flavons which contribute to the dark matter genesis. The range of $\mathcal{O}(1-10)\,\tev$ in the dark matter mass and $v_\phi$ parameter space can produce the observed relic density of the Universe.

With sufficiently large $v_\phi$, the interaction between dark matter and the Standard Model particles becomes feeble and dark matter can be produced via freeze-in mechanisms. In this case we find that the region can be divided into two parts, one dominated by the IR process such as dark matter production through pseudoscalar flavon $2\to 2$ scattering, the other is a UV process, dark matter production from the Standard Model fermion associated with a Higgs boson, $2\to 3$ scattering governed by higher dimensional operators. The choice of reheating temperature controls which scenario to dominate in the frozen-in density of the dark matter. We find for $\mathcal{O}(10)\,\gev$ pseudoscalar mediator, freeze-in production of dark matter abundance matches with the observed relic for $v_\phi$ as high as $\mathcal{O}(10^7 - 10^9)\,$GeV and dark matter in the below-TeV mass range. As the dark matter mass depends on the Froggatt-Nielsen charge of the right-handed neutrino, the suitable choices of the charges for freeze-out and freeze-in cases separately allow us to obtain the mass and mixing textures of the light neutrinos.

For the freeze-out scenario, we explore the possibility of probing the viable parameter space at direct detection experiments. Both spin-independent and spin-dependent contributions are generated in our setup. In the spin-independent case, a tree-level scalar flavon exchange and box diagram with two pseudoscalar fields give comparable contributions. The allowed cross sections fall mostly in the neutrino-fog region of the nucleon recoil experiments. The spin-dependent case is mediated by a tree-level pseudoscalar flavon, which generates cross sections that are five orders of magnitude lower than the current experimental sensitivity.

The absence of gauge interaction in this minimal setup weakens the potential of searching for dark matter through indirect detection experiments. However, the dark matter annihilation to flavon particles and its subsequent decay to SM fermions can give rise to signals in these experiments. We compare the limits obtained in~\cite{Profumo:2017obk} from the Fermi-LAT and H.E.S.S. data on the annihilation cross section with our region of interest. The parameter space satisfying the observed relic density and also allowed by bounds from other low-energy data is not constrained by these indirect detection limits.
In an attempt at UV completions, evidence at the dark matter indirect detection experiments may be explored further. In the case of less than $\mathcal{O}(\gev)$ flavons, the flavor-changing decays of $B$-meson and kaon may provide interesting signatures, which could be subjects of future investigation.

\subsection*{Acknowledgements}
The authors thank Graham Kribs for useful discussions. The work of T.T. is supported by the Universit\"at Siegen under the Young Investigator Research Group grant. R.M. acknowledges the support of the Deutsche Forschungsgemeinschaft (DFG, German Research Foundation) under Grant 396021762 - TRR 257 via Mercator Fellowship during the visit at Universit\"at Siegen. T.T. is grateful to the Mainz Institute for Theoretical Physics (MITP) of the DFG Cluster of Excellence PRISMA+ (Project ID 39083149) for its hospitality and its partial support during the completion of this work.

\appendix

\section{An example of benchmark parameters}
\label{app:yukawa}

Here we quote values of the $\mathcal{O}(1)$ numbers $c_x^{ij}$ introduced in the Lagrangian in Eq.~\ref{eq:Lag} and used in the numerical analysis of this work. 
\begin{align}
c_u = 
\left(\begin{array}{ccc}
3.0 & 1.0 & 1.0 \\
1.0 & 1.6 & 1.0 \\
1.0 & 1.0 & 1.0
\end{array}\right),\qquad 
c_d = 
\left(\begin{array}{ccc}
1.6 & 1.0 & 1.0 \\
1.0 & 1.5 & 1.0 \\
1.0 & 1.0 & 2.0
\end{array}\right), \qquad
c_e = 
\left(\begin{array}{ccc}
3.0 & 1.0 & 1.0 \\
1.0 & 1.4 & 0.3 \\
1.0 & 1.0 & 0.8
\end{array}\right).
\end{align}
These entries schematically generate the SM fermion masses and mixing textures. The SM fermion masses we obtain are listed below.
\begin{align}
&m_u\approx 3\,\text{MeV},~m_c\approx 1.23\,\text{GeV},~m_t\approx 179\,\text{GeV}, \nn \\
&m_d\approx 5\,\text{MeV},~m_s\approx 100\,\text{MeV},~m_b\approx 4.8\,\text{GeV}, \nn \\
&m_e\approx 0.5\,\text{MeV},~m_\mu\approx 105\,\text{MeV},~m_\tau\approx 1.78\,\text{GeV}. \nn
\end{align}

\section{Amplitudes and cross-sections}
\label{app:Xsec}
In this Appendix, we provide the relevant expressions for the amplitudes and cross sections used in the calculation of our results on DM genesis. 
The squared amplitude of the $N^1 N^1 \to a \, a$ process in the center-of-mass frame is given by
\begin{align*}
|\mathcal{T}|^2 = \quad &\frac{1}{v_{\phi}^4 \left(s-m_s^2\right)^2 \left(t-m_{\text{DM}}^2\right)^2 \left(u-m_{\text{DM}}^2\right)^2} \ \times \nn\\[1ex]
\Bigg\{ &2 q_{N_1}^4 m_{\text{DM}}^4 \left(s-m_s^2\right)^2 \Big[m_a^4 (t+u)^2-m_a^2 s (t-u)^2-m_a^2 (t+u)^3-t u \left(s^2-2t^2-2u^2\right)\Big] \nn\\[1ex]
+\,&2 q_{N_1}^4 m_{\text{DM}}^8 \left(s-m_s^2\right)^2 \Big[4 m_a^4-12 m_a^2 (t+u)-s^2+6 (t+u)^2\Big]
\end{align*}
\begin{align}
\quad -\,&2 q_{N_1}^4 m_{\text{DM}}^6 \left(s-m_s^2\right)^2 (t+u) \Big[4 m_a^4-6 m_a^2 (t+u)-s^2+2 (t+u)^2\Big] \nn\\[1ex]
+\,&16 q_{N_1}^4 m_{\text{DM}}^{10} \left(s-m_s^2\right)^2 \left(m_a^2-t-u\right) \nn\\[1ex]
-\,&8 \lambda_{\phi} v_{\phi}^2 \, q_{N_1}^3 m_{\text{DM}}^4 \left(s-m_s^2\right) (t-u)^2 \left(t-m_{\text{DM}}^2\right) \left(u-m_{\text{DM}}^2\right) \nn\\[1ex]
+\,&8 \lambda_{\phi}^2 v_{\phi}^4 \, q_{N_1}^2 m_{\text{DM}}^2 \left(s-4 m_{\text{DM}}^2\right) \left(t-m_{\text{DM}}^2\right)^2 \left(u-m_{\text{DM}}^2\right)^2 + 8 q_{N_1}^4 m_{\text{DM}}^{12} \left(s-m_s^2\right)^2 \Bigg\} \,,
\label{eq:AmpNNaa}
\end{align}

where $s$, $t$, and $u$ are the Mandelstam variables, $s = \ecm^2$. \\

The combined cross section of the $t$- and $u$- channels of the $N^1 N^1 \to a \, a$ process in the center-of-mass frame is given by
\begin{align}
\label{eq:XsecNNaa}
\sigma_{NN \to aa}^{t+u}(\ecm) = \frac{q_{N_1}^4 \mdm^4}{4\pi v_\phi^4 \left(\ecm^2 - 4\mdm^2\right)} \left( \tanh^{-1}\sqrt{1 - \frac{4\mdm^2}{E_\text{CM}^2}} - \sqrt{1 - \frac{4\mdm^2}{E_\text{CM}^2}} \right) \,,
\end{align}
where we treat the pseudoscalar as massless.\\

The total cross section for the $N^1 N^1 \to a \, a$ process with a massive pseudoscalar in the center-of-mass frame is given by
\begin{align}
\label{eq:XsecNNaaFull}
&\sigma_{NN \to aa}^{tot}(E_\text{CM}) \ = \ 
\frac{\lambda_{\phi}^2 q_{N_1}^2 m_{\text{DM}}^2 \sqrt{\left(\ecm^2-4 m_a^2\right) \left(\ecm^2-4 m_{\text{DM}}^2\right)}}{4 \pi \ecm^2 \left(\ecm^2-m_s^2\right)^2} \nn\\[3ex]
&+\frac{\lambda_{\phi} q_{N_1}^3 m_{\text{DM}}^4}{\pi v_{\phi}^2 \ecm^2 \left(\ecm^2-4 m_{\text{DM}}^2\right) \left(\ecm^2-m_s^2\right)} \Biggl\{ \sqrt{\left(\ecm^2-4 m_a^2\right) \left(\ecm^2-4
m_{\text{DM}}^2\right)} \nn\\[3ex]
&+\left(\ecm^2-2 m_a^2\right) \coth^{-1}\frac{2 m_a^2-\ecm^2}{\sqrt{\left(\ecm^2-4 m_a^2\right) \left(\ecm^2-4
m_{\text{DM}}^2\right)}} \Biggr\} \nn\\[3ex]
&-\frac{q_{N_1}^4 m_{\text{DM}}^4}{8 \pi v_{\phi}^4 \ecm^2 \left(\ecm^2-4 m_{\text{DM}}^2\right)} \Biggl\{ \sqrt{\left(\ecm^2-4 m_a^2\right) \left(\ecm^2-4 m_{\text{DM}}^2\right)} \left(2+\frac{ m_a^4}{\ecm^2 m_{\text{DM}}^2-4 m_a^2 m_{\text{DM}}^2+m_a^4}\right) \nn\\[3ex]
&+2 \left(\ecm^2-2 m_a^2+\frac{2 m_a^4}{\ecm^2-2 m_a^2}\right) \coth^{-1}\frac{2 m_a^2-\ecm^2}{\sqrt{\left(\ecm^2-4 m_a^2\right) \left(\ecm^2-4 m_{\text{DM}}^2\right)}} \Biggr\} \,.
\end{align}

\end{document}